 \documentclass[10pt,preprint]{aastex}

 \usepackage{longtable}
 
\newcommand{\Htwo}{H$_2$}            
\newcommand {\HI}     {\ion{H}{1}}      

\newcommand {\CI}       {\ion{C}{1}}     
\newcommand {\CII}      {\ion{C}{2}}     

\newcommand {\NI}        {\ion{N}{1}}     

\newcommand {\OI}       {\ion{O}{1}}       
\newcommand {\OVI}     {\ion{O}{6}}      

\newcommand {\PII}       {\ion{P}{2}}
\newcommand {\SiII}      {\ion{Si}{2}}

\newcommand {\ArI}       {\ion{Ar}{1}}    
\newcommand {\FeII}    {\ion{Fe}{2}}

\newcommand{\Lya}{Ly$\alpha$}
\newcommand{\Lyb}{Ly$\beta$}

\newcommand{\FUSE}{{\it FUSE}}
\newcommand{\IUE} {{\it IUE} }  
\newcommand{\FUSElong}{{\it Far Ultraviolet Spectrographic Explorer} }

\newcommand{\kms}{km~s{$^{-1}$}}

\newcommand\etal{et~al.}
\newcommand{\cd}{cm$^{-2}$}

\begin{document}

\title{A Far Ultraviolet Spectroscopic Explorer Survey of  \\ 
            Interstellar Molecular Hydrogen in the Galactic Disk}

\author{J. Michael Shull, Charles W. Danforth} 

\affil{Department of Astrophysical and Planetary Sciences and CASA \\
University of Colorado, 389-UCB, Boulder, CO 80309}

\author{Katherine L. Anderson}
\affil{Green Mountain High School, Lakewood, CO 80228, USA}

\email{michael.shull@colorado.edu, charlesdanforth@gmail.com, \\
 kalander@jeffcoschools.us}


\begin{abstract}

We report results from a \FUSElong (\FUSE) survey of interstellar molecular hydrogen 
(H$_2$) in the Galactic disk toward 139 O-type and early B-type stars at Galactic latitudes 
$|b| \leq 10^{\circ}$, with updated photometric and parallax distances.  
The H$_2$ absorption is measured using the far-ultraviolet Lyman and Werner bands,
including strong R(0), R(1), and P(1) lines from rotational levels $J = 0$ and $J = 1$
and excited states up to $J = 5$ (sometimes $J = 6$ and 7).  For each sight line, we report
column densities $N_{\rm H2}$, $N_{\rm HI}$, $N(J)$, $N_{\rm H} = N_{\rm HI} + 2N_{\rm H2}$, 
and molecular fraction, $f_{\rm H2} = 2N_{\rm H2} / N_{\rm H}$.    Our survey extends the 1977
{\it Copernicus} \Htwo\ survey up to $N_{\rm H} \approx 5 \times 10^{21}$ cm$^{-2}$.   
The lowest rotational states have excitation temperatures and rms dispersions, 
$\langle T_{01} \rangle = 88 \pm 20$~K and $\langle T_{02} \rangle = 77 \pm 18$~K, 
suggesting that $J = 0,1,2$ are coupled to the gas kinetic temperature.  Populations of 
higher-$J$ states exhibit mean excitation temperatures, 
$\langle T_{24} \rangle = 237 \pm 91$~K and $\langle T_{35} \rangle =  304 \pm 108$~K,
produced primarily by UV radiative pumping.  Correlations of $f_{\rm H2}$ with $E(B-V)$ 
and $N_{\rm H}$ show a transition to $f_{\rm H2} \geq 0.1$ at $N_{\rm H} \gtrsim 10^{21}$~\cd\ 
and $E(B-V) \gtrsim 0.2$, interpreted with an analytic model of \Htwo\ formation-dissociation 
equilibrium and attenuation of the far-UV radiation field by self-shielding and dust opacity.  
Results of this disk survey are compared to previous \FUSE\ studies of \Htwo\ in translucent 
clouds, at high Galactic latitudes, and in the Magellanic Clouds.  Using updated distances 
to the target stars, we find average sightline values $\langle f_{\rm H2} \rangle = 0.20$ and 
$\langle N_{\rm H} / E(B-V) \rangle = 6.07 \times 10^{21}~{\rm cm}^{-2}~{\rm mag}^{-1}$.  

\end{abstract}


\section{Introduction}

Molecular hydrogen (\Htwo) is the most abundant molecule in the universe, constituting 
the majority of the interstellar molecular clouds that eventually form stars.  Even though 
\Htwo\ plays an important role in the chemistry of the interstellar medium (ISM), many 
questions remain about its distribution, formation, and destruction in both diffuse and 
protostellar clouds.  The {\it Copernicus} satellite in the 1970s provided the first large-scale 
survey of interstellar \Htwo\ (Spitzer \etal\ 1974;  Savage \etal\ 1977).  With the 1999 launch 
of the \FUSElong (\FUSE) satellite, astronomers once again gained access to the 
far-ultraviolet (FUV) wavelengths needed to study \Htwo\ in its resonance absorption lines.
This survey extends these studies to 139 OB-type stars, many of them fainter and more 
distant  than observed by {\it Copernicus}.  

From the inception of the \FUSE\ mission, \Htwo\ studies were part of the science plan.  
The \FUSE\ satellite, its mission, and its on-orbit performance were described in 
Moos \etal\ (2000) and Sahnow \etal\ (2000).  Initial FUSE studies of \Htwo\ were 
reported in papers from the Early Release Observations (Snow \etal\ 2000; Shull \etal\ 
2000).  Later studies included a survey of 70 sight lines to OB stars probing \Htwo\ in the 
Large and Small Magellanic Clouds (Tumlinson \etal\ 2002), surveys of 38 sight lines 
through translucent clouds (Rachford \etal\ 2002, 2009), studies of chemical relationships
 of \Htwo\ with other molecules (CO, CH, CH$^+$, CN) and atomic species 
(Burgh \etal\ 2007; Sheffer \etal\ 2008; Jensen \etal\ 2010), and observations of \Htwo\ 
in the low Galactic halo (Gillmon \etal\ 2006;  Wakker 2006; Gillmon \& Shull 2006).  
At high redshift, \Htwo\ Lyman/Werner lines have been detected in damped \Lya\ 
absorbers (Noterdaeme \etal\ 2008; Jorgenson \etal\ 2014; Balashev \etal\ 2019).

We present the results of a \FUSE\ survey of interstellar \Htwo\ absorption in the Galactic disk, 
using transitions from the ground electronic state, $X\,^{1} \Sigma_g^+$, to excited electronic 
states, $B\,^{1}\Sigma_u^+$ (Lyman bands) and $C\,^{1}\Pi_u$ (Werner bands).  These 
lines are rovibrational transitions from lower states ($v_l, J_l$) to upper states ($v_u, J_u$).  
In the  cold, low-density ISM, essentially all the molecules are in the ground vibrational state 
($v_l = 0$).   Absorption lines were observed up to $J_l = 5-6$ and occasionally $J_l = 7$.  
Because the wavefunction for the homonuclear \Htwo\  is anti-symmetric under 
 interchange of the identical (fermionic) protons, the even-parity rotational states ($J = 0, 2, 4...$) 
 have total nuclear spin $S = 0$ (spin anti-symmetric para-\Htwo) while odd-parity states 
 ($J = 1, 3, 5...$) have $S = 1$ (spin-symmetric ortho-\Htwo).  
The statistical weights of these states are $g_J = (2S+1)(2J+1)$, with odd-$J$ states having a 
factor of three higher weight.   Absorption bands leading to the upper level are identified  
by changes in vibrational state ($v_u - v_l)$ and rotational angular momentum ($J_u - J_l$).  
The upper electronic state ($^1\Sigma_u^+$) of the Lyman bands has angular momentum 
$\Lambda = 0$ along the internuclear axis.  Dipole-allowed changes in rotational state 
($\Delta J = \pm 1$) are denoted as R-branch ($J_u = J_l + 1$) and P-branch ($J_u = J_l - 1$).    
The upper electronic state ($^1\Pi_u$) of Werner bands has $\Lambda = 1$, allowing a 
Q-branch ($J_u = J_l$) in addition to the R and P branches.     

We observed multiple lines in the Lyman and Werner bands in the FUV (930--1126 \AA)  
towards background OB stars located near the Galactic disk plane at $|b| \leq 10^{\circ}$.  
Several target stars were outside this latitude range, but all show strong \Htwo\ absorption.  
Lines from $J =0$ and $J = 1$ nearly always exhibit damping wings.  
Analysis of absorption-line equivalent widths and damping wings yields column 
densities, $N(J)$, in individual rotational states.  {\bf Figure 1} shows a \FUSE\ spectrum of the 
sight line to HD~46150, an O5~Vf star at 1.5~kpc distance and $E(B-V) = 0.45$, with the Lyman 
and Werner bands labeled including a close-up of the (4-0) Lyman band.    The total \Htwo\ 
column density, $N_{\rm H2}$, is found by summing over all observed $J$ states.  Typically,  98\% 
to 99\% of the molecules reside in the lowest two levels, $J = 0$ and $J = 1$.  

Observations of \Htwo\ column densities, the molecular fraction in diffuse clouds, and its 
rotational excitation provide diagnostics of diffuse ISM (Shull \& Beckwith 1982). The excitation 
temperature, $T_{01}$, of the lowest rotational states (para-\Htwo\ in $J = 0$ and ortho-\Htwo\ 
in $J = 1$) is an approximate measure of the gas kinetic temperature.   In many cases, the three 
lowest levels ($J = 0, 1, 2$) appear to be thermalized, with both $T_{01}$ and $T_{02}$ coupled 
to the kinetic temperature (Gry \etal\ 2002; Le Petit \etal\ 2006).   In diffuse interstellar clouds, 
thermal equilibration requires sufficiently large gas densities for proton interchange collisions 
to produce ortho-para conversion between $J = 0$ and $J =1$ (Dalgarno \etal\ 1973; Gerlich 1990).  
Column densities $N_{\rm H2} \geq 10^{18}$~\cd\ are needed to produce strong self-shielding in 
absorption lines from the dissociating FUV radiation.  The current observations show that both 
$T_{01}$ and $T_{02}$ provide estimates of the heating and cooling processes in the diffuse ISM.  
The excitation temperature, $T_{\rm exc}$, of higher rotational states ($J = 3-6$) is influenced 
by the FUV radiation field, which excites and photodissociates \Htwo\ (Jura 1974; Black \& Dalgarno 
1976).  From \FUSE\ data on these excited states, we compute temperatures, $T_{24}$ and 
$T_{35}$, based on population ratios of $J = 2$ to 4 and $J = 3$ to 5, respectively.

Section~2 begins with a description of the OB-star sample and our methods of 
data acquisition and analysis.    Section~3 provides the survey results, including \Htwo\ and \HI\ 
column densities, molecular fractions, rotational excitation temperatures, and gas-to-dust ratio, 
$N_{\rm H} / E(B-V)$.  We also present a simple analytic model of the transition from \HI\ to \Htwo,
including \Htwo\ formation rates, FUV dissociation, self-shielding, and dust opacity.  The 
transition occurs at $f_{\rm H2} \approx 0.1$ at dust optical depth $\tau_d \approx 1$ and is 
controlled primarily by the ratio of FUV flux to gas density.  The lowest rotational states 
exhibit mean excitation temperatures $\langle T_{01} \rangle = 88 \pm 20$~K 
and $\langle T_{02} \rangle = 77 \pm 18$~K.   Higher rotational states ($J \geq 3$) have larger 
excitation temperatures, $T_{\rm exc} \approx 150-650$~K, arising from fluorescent cascade 
following FUV radiative excitation in the Lyman and Werner bands.   Section~4 summarizes 
our findings, with comparisons to the {\it Copernicus} survey of \Htwo\ (Savage \etal\ 1977) 
and the $N_{\rm H}/E(B-V)$ ratio (Bohlin \etal\ 1978).   The \FUSE\ mean value for the Galactic disk,
$\langle N_{\rm H} / E(B-V) \rangle = (6.07\pm1.01) \times 10^{21}$ cm$^{-2}$~mag$^{-1}$, is 
lower than estimates from 21-cm/far-IR studies  at high Galactic latitudes (Liszt 2014a,b), 
suggesting that gas and dust have different spatial distributions above the disk.


\section{Data Acquisition and Reduction}

\subsection{FUSE Observations}

The 139 targets in this survey were drawn primarily from \FUSE\ programs designed to study 
OB stars and interstellar gas in the Milky Way.  {\bf Figure 2}  shows the target distribution in
Galactic longitude and latitude, with color coding for their distances.  Most stars are O-type and
early B-type (B0, B0.5) and generally more distant than those in the 1977 {\it Copernicus} 
survey.   Many were part of  \FUSE\ science team projects to study \OVI\ from hot gas in the 
Galactic disk (Bowen \etal\ 2008), interstellar D/H (Moos \etal\ 2002; Hoopes \etal\ 2003; 
H\'ebrard \etal\ 2005), and hot stars and their winds (Massa \etal\  2003).  Several stars were 
analyzed specifically for their interstellar \Htwo\ (Shull \etal\ 2000;  Snow \etal\ 2000).  

{\bf Table 1} lists information on the 139 stellar targets and their observational characteristics.  
These stars have updated spectral types (SpT) and both photometric and {\it Gaia}-DR2 parallax 
distances determined by Shull \& Danforth (2019) using new information from the Galactic O-star 
Spectroscopic Survey (Ma\'iz Apell\'aniz \etal\ 2004).  The GOS project generated a large sample of
O~stars within several kiloparsecs of the Sun with updated spectral classification (Sota \etal\ 2011, 
2014), together with digital photometry and optical-NIR dust extinction 
(Ma\'iz-Apell\'aniz \& Barb\'a 2018).    The first ten columns of Table 1 provide our internal target ID, 
star name, Galactic coordinates ($\ell$, $b$), photometry ($B$ and $V$ magnitudes), color excess 
$E(B-V)$, SpT, and both photometric and parallax distances.  The last two columns list  the \FUSE\ 
program ID and exposure time of the primary observation.  In some cases, we used other \FUSE\ 
observations to supplement or confirm our measurements.   Footnotes explain the sources for 
photometry and distances.  

All column densities are expressed in units of \cd\ and  often quoted in logarithmic format ($\log N$).  
The total hydrogen column density, $N_{\rm H} = N_{\rm HI} + 2N_{\rm H2}$, is used to derive the 
molecular fraction, $f_{\rm H2} = 2 N_{\rm H2} / N_{\rm H}$, and explore its correlations with $E(B-V)$ 
and line-of-sight pathlength.  Most \HI\ column densities, $N_{\rm HI}$, were taken from previous 
 \Lya\ profile-fitting surveys  (Shull \& Van Steenberg 1985; Diplas \& Savage 1994; Jenkins 2019)
supplemented by a few individual \Lya\ measurements.   {\bf Table 2} presents our adopted values 
of $N_{\rm HI}$ and a comparison to previous measurements.  We adopted $N_{\rm HI}$ in priority 
order of:  (J19) 57 stars from Jenkins (2019);  (DS94) 51 stars from Diplas \& Savage (1994);  
(FM90) two stars from Fitzpatrick \& Massa (1990); and three stars from other sources.  A comparison 
between J19 and DS94 found generally good agreement, within $0.03-0.08$ in $\log N$.  Only
seven sight lines differed by larger amounts ($0.09-0.19$). For 26 sight lines with no available no 
\Lya\ fits, we estimated $N_{\rm HI}$ from the scaling relation (Bohlin \etal\ 1978) of total hydrogen 
column density with color excess, 
$N_{\rm H} = (5.8 \times 10^{21}~{\rm cm}^{-2}~{\rm mag}^{-1}) E(B-V)$,  where 
$N_{\rm HI} = N_{\rm H} - 2N_{\rm H2}$.   

\subsection{Data Analysis}

The data analysis for this survey used numerous absorption lines from \Htwo\ Lyman and 
Werner transitions between  930~-~1126 \AA.  Our methodology is identical to that used in the
\FUSE\ survey of \Htwo\ in the LMC and SMC (Tumlinson \etal\ 2002), and we refer readers to 
that paper for details.  We search for absorption lines from $J = 0-7$ and typically detect 
lines up to $J=5$.  Most lines exhibit shifts in wavelength (0.03--0.13~\AA) with offsets 
determined by measuring narrow interstellar metal lines such as \ArI (1048.218~\AA, 
1066.660~\AA),  \FeII\ (1055.269~\AA, 1063.177~\AA, 1144.938~\AA), \SiII\ (1020.699~\AA), 
\PII\ (1152.818~\AA), and other lines of \FeII, \OI, \NI.   This procedure is similar to that employed
 by Tumlinson \etal\ (2002), Wakker (2006), and Gillmon \etal\ (2006).  We also considered the
R(0) lines of HD in its (3-0), (4-0), (5-0) bands at 1066.271~\AA, 1054.286~\AA\ and 
1042.847~\AA, respectively.   However, the HD lines were often offset in velocity from \Htwo, 
possibly because of its presence in only one of the components of the \Htwo\ line profile.   

The most accurate determinations of column densities $N$ come from weak lines where
equivalent width $W_{\lambda} \propto N$, or from strong lines with damping wings
where $W_{\lambda} \propto N^{1/2}$.  In both limits, the column density can be determined 
without knowledge of the Doppler parameter ($b$).  Between these two regimes, from 
line-center optical depth $\tau_0 \approx 1$ up to $\tau_{\rm damp} \approx 10^{3-4}$, one 
must employ curve-of-growth (CoG) methods\footnote{The equivalent width ($W_{\lambda}$) 
of a saturated absorption line rises slowly with increasing column density $N$ and line-center 
optical depth $\tau_0$.  As shown by asymptotic analysis, 
$W_{\lambda} / \lambda \approx (2b/c) [ \ln (\tau_0 / \ln 2 )]^{1/2}$ for large $\tau_0$ and a 
Doppler-broadened Gaussian velocity profile, $\phi(v) \propto  \exp (-v^2/b^2)$.  Draine (2011) 
noted that this formula is accurate to 5\% for $1.254 < \tau_0 < \tau_{\rm damp}$, up to the 
onset of damping wings.  Inverting this relation, we find
$\tau_0 \approx (0.693) \exp [ (W_{\lambda}/\lambda)^2 / (2b/c)^2 ]$.  Clearly, $\tau_0$ and 
$N$ are exponentially sensitive to measurements of $W_{\lambda}$ and $b$.  When 
equivalent widths of the accessible lines lie on a flat CoG, errors can exceed $\pm0.40$ for 
$\log N = 16-17$ and $\pm0.50$ for $\log N = 17-19$.} that require an estimation of $b$.   
The linear relation between $W_{\lambda}$ and $N$ and the onset of line saturation 
are measured by optical depth at line center,
\begin{eqnarray} 
   W_{\lambda} &=& \left( \frac {\pi e^2}{m_e c} \right) \frac {N f \lambda^2} {c} = 
         (88.53~{\rm \AA}) N_{15} \,  \lambda_{1000}  \left[  \frac { f  \lambda} {10~{\rm \AA}} \right]  \\
    \tau_0  &=& \left( \frac {\pi e^2}{m_e c} \right) \frac {N f \lambda} { \sqrt{\pi} \,  b} = 
         (1.497) N_{15}  \,  b_{10}^{-1} \left[  \frac {f  \lambda} {10~{\rm \AA}} \right] \;  .  
\end{eqnarray}
Here, we scaled to column density $N = (10^{15}~{\rm cm}^{-2})N_{15}$, Doppler parameter 
$b = (10~{\rm km~s}^{-1})b_{10}$,  and wavelength $\lambda = (1000~{\rm \AA}) \lambda_{1000}$.  
Line strengths are normalized to $f \lambda = 10$~\AA, a typical value for many lines in the 
Lyman and Werner bands.  

Equivalent widths of the highly saturated lines from $J = 2$ and $J = 3$ are frequently  quite
large, with dimensionless values $W_{\lambda} / \lambda \approx (1-3) \times10^{-4}$.   
On the flat portion of the CoG, these widths are primarily determined by the effective $b$-value, 
with asymptotic values $W_{\lambda}/\lambda \approx (2b /c) [ \ln (\tau_0 / \ln 2 )]^{1/2}$.   
As shown in the CoG suite prepared by McCandliss (2003), strong lines on the equivalent 
width plateau often require Doppler parameters, $b = 10-20$~\kms, much higher than the 
expected thermal values, 
$b_{\rm th} = (2kT/m)^{1/2} \approx (0.9~{\rm km~s}^{-1}) T_{100}^{1/2}$, 
for \Htwo\ at temperatures $T = (100~{\rm K})T_{100}$.  They likely result from velocity 
components within the absorber and micro-turbulence.  Several previous studies 
(Jenkins \& Peimbert 1997; Gry \etal\ 2002; Lacour \etal\ 2005) suggested that $b$-values 
increase with rotation level.  
Although we found several cases where the higher-$J$ lines required larger $b$-values, 
we saw no evidence of  a uniform trend in the current survey or in our study of \Htwo\ in 
the Magellanic Clouds (Tumlinson \etal\ 2002).  

For absorbers with $N(J) \leq 10^{18}$~\cd, we measured equivalent widths of the accessible 
\Htwo\ lines, producing a CoG that yields a $b$ value and column densities, $N(J)$, in rotational 
states $J = 0-5$, and occasionally $J = 6$ and $J = 7$.   {\bf Table 3} lists column densities in 
$J = 0-6$ and the inferred $b$-value when measurable. 
Most column densities range from $\log N(J)  \approx 19.0-21.5$ in $J = 0-1$ to 
$\log N(J) \approx 13.7-15.5$ in $J = 4-6$.   In 48 targets, we found detectable column densities 
in $J = 6$. For other sight lines, we quote upper limits, which range from $\log N(6) < 13.85-14.40$ 
depending on the data quality.   Six targets had detectable column densities in $J = 7$, described 
in footnote (a) to Table~3.   

For absorbers with damping wings in R(0), R(1), and P(1) lines, we fitted line profiles
to derive $N(0)$ and $N(1)$, a technique used in previous studies of \Htwo\ (Savage \etal\ 
1977;  Tumlinson  \etal\ 2002; Jenkins 2019).  
The errors on these column densities depend on data quality, both the signal-to-noise (S/N) 
and our ability to define the continuum on either side of the blended $R(0)$, $R(1)$, $P(1)$ 
complex (see the bottom panel of Figure 1 for an example).   Based on our experience with
fitting damping wings, we define three levels of S/N ratio and corresponding errors, 
$\sigma_{\log N}$, on $\log N(0)$ and $\log N(1)$:
(1) $S/N \geq 15$ (0.03--0.05 errors); 
(2) $5 \leq S/N \leq 15$ (0.05--0.10); and
(3) $S/N \leq 5$ (0.10--0.20).

A key aspect of our analysis software is the rapid and consistent measurement of as many 
individual \Htwo\ absorption lines as possible in each sight line.  We then use CoGs to derive 
column densities for the undamped lines in $J \geq 2$.  The programs are written to 
consistently measure all available (unblocked) absorption lines in the spectra.  In many 
instances, lines from $J = 2$ and $J = 3$ were strongly saturated,  with equivalent widths 
lying on the flat (Doppler-broadened) portion of the CoG where column densities are difficult to 
determine.  In this case, $W_{\lambda}$ depends primarily on the Doppler parameter, which is 
often much larger than the thermal value because of multiple velocity components.  
The process of measuring the numerous lines that enter the CoG fitting cannot be automated 
completely.  Our software requires the user to decide on a line-by-line basis which \Htwo\ lines 
will be fitted.  We ignore \Htwo\ lines in spectral regions near strong interstellar absorption or 
bright geocoronal emission.  For example, the \Htwo\ Lyman (6-0) band is obscured by strong 
damping wings of the interstellar \Lyb\ line (1025.722~\AA), and the Lyman (5-0) band lies 
among the resonance absorption lines of \CII\ (1036.337~\AA) and \CII$^*$ (1037.018~\AA).   
We neglect  these bands except when the higher-$J$ lines are separated from the 
intervening absorption.   
  
For the higher-$J$ lines, the software steps through the expected positions of the lines, 
band by band, shifted by the approximate velocity offset from the metal lines.  
Going through each complete band of lines consecutively, the routine displays the area where a 
line should appear.  If the line is present and unblended with any other transitional or metal line, 
it is fitted to a Gaussian, and its equivalent width, errors, wavelength, full width half maximum, and 
velocity offset are entered into a table.  The Lyman (6-0) band is typically omitted, and the 
Lyman (5-0) band is measured only partially, because of overlap with  \CII\ $\lambda 1036$ 
absorption.  In these bands, lines that are observable and undistorted are measured and included
in the tables.  This routine is executed for each detector segment, creating four tables of data, with 
maximum redundancy over the critical range between 1000~\AA\  and 1126~\AA.  The four tables 
are then merged, and a CoG is generated with a single Doppler parameter.  An error routine is run 
to calculate the smallest errors over a range of $b$-values.  Asymmetric error bars are generated 
for $b$ and column densities.    

As noted above, column densities from $J = 2$ and $J = 3$ are difficult to measure, because most 
of their lines are highly saturated.  To alleviate CoG uncertainties and determine an accurate 
$b$-value it is helpful to measure the weakest available lines, with strengths $f \lambda \leq 4$~\AA.   
These weak lines are useful in anchoring the CoG  for more saturated lines.   Unfortunately, we 
were unable to measure the weakest lines from $J = 2$:  (0-0) P(2) at 1112.495~\AA\
($f \lambda = 0.740$~\AA);  (0-0) R(2) at 1110.120~\AA\ ($f \lambda = 1.199$~\AA); and 
(1-0) R(2) 1094.244~\AA\ ($f \lambda = 4.016$~\AA).  These lines are blended with (0-0) P(1), 
(0-0) R(3), and (1-0) P(1), respectively.  For $J = 3$, we were unable to measure the two weakest 
lines: (0-0) P(3) at 1115.896~\AA\ ($f \lambda =  0.784$~\AA) blended with (0-0) R(4) at 
1116.013~\AA, and (0-0) R(3) at 1112.584~\AA\ ($f \lambda = 1.135$~\AA) blended with (0-0) P(2) 
at 1112.495~\AA.  The weakest line that could be measured was (1-0) P(3) at 1099.788~\AA\ 
($f \lambda = 2.639$~\AA).   For $J = 4$, the weakest lines, (0-0) P(4) at 1120.247~\AA\ 
($f \lambda = 0.808$~\AA) and (0-0) R(4) at 1116.013~\AA\ ($f \lambda = 1.116$~\AA), are 
blended with (0-0) P(3) 1115.896~\AA\ and (0-0) R(5) 1120.300~\AA, respectively.  In some cases 
we were able to separate (0-0) P(4).  In general the most accessible lines were the P(4) lines 
in (1-0), (2-0), (3-0), (4-0), (5-0), (7-0) bands and the R(4) lines in (3-0), (4-0), (5-0) bands.  
For $J = 5$, we found a sufficient number of lines with a range of strengths for
a reliable CoG. 

A systematic uncertainty in our results comes from the possibility of multiple components in 
the absorption lines caused by more than one cloud in the line of sight.   Many spectra 
($\sim$ 40 sight lines) have velocity components that make the neighboring lines visually 
identifiable, but not separable without careful profile fitting using other information 
from higher resolution optical lines.  Some absorbers in this survey have components that 
are not resolvable; those lines are treated as though they are a single component.  
The spectra of 15 targets exhibited obvious multiple components separated by 
$\Delta v \geq 20$ \kms.  We measured those individually using a double Voigt profile fit.
The components have their values entered into separate data tables, enabling us to 
generate two CoGs with individual $b$ values.   In these cases, we report the total column 
densities measured for saturated lines or lines without visible structure and individual 
column densities where components are measurable.  Lines at $J \leq 3$ are often 
saturated and too strong to show component structure.  These lines are fitted as a single 
component, and a total column density is returned.


\section{Results}

\subsection{Molecular Abundances}

The \FUSE\ survey finds \Htwo\ everywhere in the disk of the galaxy ({\bf Figure~2}) at typical 
Galactic latitudes $|b| < 4^{\circ}$.    The \Htwo\ column density rises rapidly above
$N_{\rm H2} \geq 10^{19.5}~{\rm cm}^{-2}$ for sight lines with color excess $E(B-V) \gtrsim 0.2$, 
as illustrated in {\bf Figure~3}.  The molecular fraction, $f_{\rm H2} = 2 N_{\rm H2} / N_{\rm H}$, 
quantifies the number of hydrogen nuclei bound into \Htwo\ molecules, where 
$N_{\rm H} = N_{\rm HI} + 2N_{\rm H2}$ is the total hydrogen column density.   
Later in this section, we will adopt $N_{\rm H2} = 10^{19.5}~{\rm cm}^{-2}$ and 
$f_{\rm H2} = 0.1$ as nominal values of the atomic-to-molecular transition in the absorbing 
gas.  With the exception of  sight lines toward  HD~3827 ($\log N_{\rm H2} = 17.48$), HD~201638 
($\log N_{\rm H2} = 18.23$), and HD~92554 ($\log N_{\rm H2} = 18.93$), all targets in the 
survey have $\log N_{\rm H2} \geq 19.0$.  {\bf Table~4} summarizes the column densities 
($N_{\rm HI}$, $N_{\rm H2}$, $N_{\rm H}$) together with the molecular fractions and rotational 
excitation temperatures inferred from populations in levels $J = 0-5$.   Rotational excitation 
temperatures, $T_{01}$, $T_{02}$, $T_{24}$, and $T_{35}$, are discussed further in 
Sections~3.4 and 3.5.  

The atomic-to-molecular transition arises from  both \Htwo\ self-shielding in optical thick lines 
and dust attenuation of  the dissociating FUV radiation (Browning \etal\ 2003; Krumholz \etal\ 
2008, 2009; Sternberg \etal\ 2014).  The {\it Copernicus} survey of 61 stellar targets 
(Savage \etal\ 1977) noted that $f_{\rm H2}$ rises rapidly above 1\% at $E(B-V) \geq 0.08$ 
and $N_{\rm H} \geq 5 \times10^{20}$~\cd.  A transition to $f_{\rm H2} = 0.1$ appeared at lower 
hydrogen column densities ($\log N_{\rm H} \approx 20.4$) in sight lines at high Galactic latitude 
(Gillmon \etal\ 2006).   In the lower-metallilcity LMC and SMC (Tumlinson \etal\ 2002; Browning 
\etal\ 2003) the transition shifted to higher column densities ($\log N_{\rm H} \approx 21.3-22.0$), 
consistent with models with lower metallicity but a larger ratio of FUV radiation to gas density.  
A lower grain abundance, owing to fewer refractory heavy elements, will reduce the \Htwo\
formation rate.  It also allows deeper penetration of FUV photons into the cloud, resulting in 
more \Htwo\ dissociation.   These effects are discussed further in Section 3.3.  

{\bf Figure~4} presents the distributions of molecular fraction vs.\ $E(B-V)$ and $N_{\rm H}$ 
for 139 targets in the \FUSE\ survey.  {\bf Figure 5} plots mean values of $f_{\rm H2}$ and 
hydrogen density $n_{\rm H} = N_{\rm H}/D$ along the sight lines vs.\ target distance $D$.  
These averages were evaluated as
$\langle n_{\rm H} \rangle = \sum N_{\rm H} / \sum D$ and 
$\langle f_{\rm H2} \rangle = \sum 2N_{\rm H2} / \sum [N_{\rm HI} + 2 N_{\rm H2}]$.
Both parallax and photometric distance are shown in the figure, with error bars 
reflecting formal parallax uncertainties from {\it Gaia}-DR2.  
With many sight lines to bright OB stars ($D \leq 2$~kpc), the {\it Copernicus} survey 
showed an increase of $f_{\rm H2} \geq 0.01$ at $E(B-V) \geq 0.08$ (Savage \etal\ 1977).
The  \FUSE\ survey includes more distant sight lines, with column densities up to 
$\log N_{\rm H} \approx 21.65$.  We found a transition at $f_{\rm H2} \geq 0.1$ 
at $E(B-V) \gtrsim 0.2$, $N_{\rm H} \gtrsim 10^{21}$~\cd\ and 
$N_{\rm H2} \gtrsim 10^{19.5}$~\cd.  The molecular fractions rise from 
$f_{H2} \approx 1$\% up to 40-75\% in translucent sight lines with 
$\log N_{\rm H} \approx 21.40$ to 21.65.  Using new photometric distances toward 
129 stars at $D_{\rm phot}  \leq 5$~kpc, we find a sightline-averaged hydrogen density
$\langle n_{\rm H} \rangle = 0.50$~cm$^{-3}$ and molecular fraction 
$\langle f_{\rm H2} \rangle = 0.20$.  These mean values shift toward higher values 
({\bf Table 5}) in sub-samples with $D_{\rm phot} \leq 2$~kpc, owing to bias in
distant targets that avoid heavily reddened gas.  For stars at $D \leq 2$~kpc, 
$\langle n_{\rm H} \rangle = 0.81$~cm$^{-3}$ and molecular fraction 
$\langle f_{\rm H2} \rangle = 0.27$.

\subsection{Hydrogen Gas vs.\  Extinction: $N_{\rm H}/E(B-V)$ Ratio }

The optical extinction along Galactic  sight lines is often taken to be proportional to the 
dust column density, and therefore to the gas column density.  This ``gas-to-dust ratio" assumes 
a homogeneous mixture of interstellar hydrogen and grains, which may be a good assumption
for most regions of the diffuse ISM.   Deviations can be produced by changes in 
the grain size distribution and other physical properties that arise within dark clouds such 
as $\rho$~Oph (Bohlin \etal\ 1978; Green \etal\ 1992) or in regions where shock waves have 
sputtered or destroyed some of the grains (Seab \& Shull 1983).  The total hydrogen column 
density, $N_{\rm H} = N_{\rm HI} + 2 N_{\rm H2}$, is often compared to dust content through its 
ratio to color excess, $N_{\rm H}/E(B-V)$, derived from UV surveys of \HI\ (\Lya) and \Htwo\ 
toward early-type stars (Bohlin \etal\ 1978; Savage \etal\ 1977).

Our \FUSE\ survey of the Milky Way disk should be more robust, with more stellar sight lines,
updated O-star photometry and SpTs from the GOS survey, and newly derived values of 
$E(B-V)$ and target distances (Shull \& Danforth 2019).  {\bf Figure~6} shows the distribution 
of  $N_H/E(B-V)$ vs.\ $E(B-V)$.   For 129 stars at $D \leq 5$~kpc, we find a mean ratio, 
$\langle N_{\rm H} / E(B-V) \rangle = (6.07\pm1.01) \times 10^{21}~{\rm cm}^{-2}~{\rm mag}^{-1}$
with (rms) variations shown as blue wash.  A sub-sample of 56 stars at $D \leq 2$~kpc (Table 5)
has $\langle N_{\rm H}/E(B-V) \rangle = 6.00 \times 10^{21}~{\rm cm}^{-2}~{\rm mag}^{-1}$.
Both values are slightly above the values $5.8 \times 10^{21}~{\rm cm}^{-2}~{\rm mag}^{-1}$ 
in the {\it Copernicus} survey of 75 stars (Bohlin \etal\ 1978) and 
$5.94 \times 10^{21}~{\rm cm}^{-2}~{\rm mag}^{-1}$ in the  \FUSE\ survey of 38 translucent 
sight lines with $A_V \approx 0.5-4.7$ (Rachford \etal\ 2009).  
 
Recent studies find a larger ratio when $N_{\rm HI}$ is measured from 21-cm emission and 
$E(B-V)$ is inferred from all-sky maps (Schlegel \etal\ 1998;  Schlafly \& Finkbeiner 2011) 
of  far-infrared (FIR) dust emission from {\it IRAS} and {\it COBE}/DIRBE.   Liszt (2014a,b) 
found $N_{\rm HI}/E(B-V) = 8.3 \times 10^{21}~{\rm cm}^{-2}~{\rm mag}^{-1}$ in high-latitude 
sight lines ($|b| > 20^{\circ}$) with low extinction, $0.015 < E(B-V) < 0.075$.   Lenz \etal\ (2017)
found a similar large value, $8.8 \times 10^{21}~{\rm cm}^{-2}~{\rm mag}^{-1}$.  These studies 
used only \HI, but as the authors comment, corrections for \Htwo\ are normally small for 
$E(B-V) < 0.08$.  

We investigated whether sight lines with low $E(B-V)$ have different dust-to-gas ratios in
the UV surveys.   For 25 stars with $0.01 \leq E(B-V) \leq 0.08$ in the {\it Copernicus}/\IUE\ 
survey (Bohlin \etal\ 1978) the mean ratio is $4.0\times10^{21}~{\rm cm}^{-2}~{\rm mag}^{-1}$.  
For the 21 stars in our \FUSE\ survey with $E(B-V) \leq 0.25$, we find a mean ratio 
$N_{\rm H}/E(B-V) = 5.83 \times 10^{21}~{\rm cm}^{-2}~{\rm mag}^{-1}$.  We omitted one 
outlier (HD~3827) with $\log N_{\rm H} = 20.55$, with an uncertain $E(B-V) \approx 0.02$, 
and a high ratio $N_{\rm H}/E(B-V) \approx 17 \times 10^{21}~{\rm cm}^{-2}~{\rm mag}^{-1}$.  
Located at photometric distance $D_{\rm phot} \approx 1.88$~kpc, HD~3827 lies 
$700-800$~pc below the Galactic plane at $b = -23.21^{\circ}$.  Its color excess,
$E(B-V) = 0.02$, was based on magnitudes $B = 7.76$ and $V = 8.01$ 
(Deutschman \etal\ 1976) and intrinsic color $(B-V)_0 = -0.27$.  Jenkins (2019) listed
$E(B-V) = 0.05$ for this star, based on $B = 7.76$ and $V = 7.95$,  which would 
reduce the ratio to $7 \times 10^{21}~{\rm cm}^{-2}~{\rm mag}^{-1}$.  

The difference between the two techniques (UV absorption and radio/FIR emission) 
appears to be an effect only seen at high Galactic latitudes (Liszt 2014b; Hensley \& 
Draine 2021).  Elevated ratios from UV data do not appear toward disk stars with low 
$E(B-V)$.   We suggest that the high ratios in 21-cm/FIR  measurements result from 
different distributions of gas and dust above the disk.  Dust grains are produced by 
stars in the disk and grow in the ISM through accretion of refractory elements.  Some
grains are transported above the disk plane by radiation pressure and supernova-driven 
outflows.  Other grains may settle gravitationally into lower scale-height distributions, 
separating from the high-latitude \HI.   Dust at high latitudes may also come into contact 
with hot coronal gas at $10^{6-7}$~K and experience erosion by thermal sputtering and 
destruction by fast interstellar shock waves (Jones \etal\ 1996; Slavin \etal\ 2004).  
In hot, low-density halo gas, grain lifetimes from sputtering are 
$t_{\rm sp} \approx (1~{\rm Gyr})(10^{-3}~{\rm cm}^{-3}/n_e)$.  Thus,  the 21-cm and 
far-IR surveys at $|b|  > 20^{\circ}$ likely probe systematically lower dust-to-gas ratios.

\subsection{Atomic to Molecular Transition}

As we will describe, the molecular transition from \HI\ to \Htwo\ is consistent with models 
involving \Htwo\ self-shielding and efficient \Htwo\ formation by atomic processes on
grain surfaces (Hollenbach \etal\ 1971; Jura 1975a,b; Shull \& Beckwith 1982). In equilibrium, 
the abundance ratio, $n_{\rm H2}/n_{\rm H}$, can be expressed as a balance between \Htwo\ 
formation and photo-dissociation.  For number densities $n_{\rm HI}$,  $n_{\rm H2}$, 
and total hydrogen $n_{\rm H}$, molecule formation occurs at a rate per unit volume, 
$R n_{\rm H} n_{\rm HI}$.  The total hydrogen density serves as a proxy for dust grains, whose 
surfaces are catalysts for \Htwo\ formation.  The coefficient $R$ depends on the dust-to-gas 
ratio, metallicity, and atom-grain collisional rates, and it likely varies with gas temperature, 
grain temperature, and surface physics (Hollenbach \& McKee 1979).  

Previous studies (Browning \etal\ 2003;  Krumholz \etal\ 2008, 2009; Sternberg \etal\ 2014;
Bialy \& Sternberg 2016) analyzed the molecular transition with radiative transfer.  Here, we 
present a simple analytic description of the transition (at $f_{\rm H2} \approx 0.1$) tied directly 
to parameters that control the \Htwo\ equilibrium between formation and destruction and the 
attenuation of  FUV flux by \Htwo\ self-shielding and dust opacity.  In equilibrium, 
\begin{equation}
      n_{\rm HI} \, n_{\rm H} \, R =  (f_{\rm diss} G \beta_0) n_{\rm H2} \, S_{\rm H2} \, e^{-\tau_d}   \; ,
\end{equation}
where $\beta_0$ is the average unshielded absorption rate of \Htwo\ in the Lyman and 
Werner bands.  For the local ISM,  Jura (1974) estimated 
$\beta_0 = 5 \times 10^{-10}~{\rm s}^{-1}$ with $f_{\rm diss} \approx 0.11$.  With updated 
line-dissociation data and models of self-shielding, Draine \& Bertoldi (1996) found a mean 
fraction $\langle f_{\rm diss} \rangle \approx 0.15$ for all \Htwo\ absorptions inside the cloud.
The parameter $G$ allows for local elevation of the FUV radiation relative to its average value 
(Habing 1968), $\tau_d$ is the dust optical depth at 930--1130~\AA, and the factor 
$S_{\rm H2}$ accounts for \Htwo\ self-shielding as the absorption lines become optically thick.  
For clarity, we define the {\it local} molecular fraction $f  = 2n_{\rm H2}/n_{\rm H}$, with 
$n_{\rm HI} = (1-f)n_{\rm H}$.  This fraction varies with depth into the cloud, whereas the 
{\it observed} fraction, $f_{\rm H2} \equiv 2N_{\rm H2}/N_{\rm H}$, depends on the integrated 
column densities.  This leads to an expression,
\begin{equation}
      \frac { f }  {(1-f)} =  \left( \frac  {2n_{\rm H} R}  {f_{\rm diss} \beta_0 \, G} \right) 
                    S_{\rm H2} ^{-1} \, e^{\tau_d}  \;  .
 \end{equation}
Draine \& Bertoldi (1996) provided a reliable approximation,
 $S_{\rm H2} = [ N_{\rm H2} / 10^{14}~{\rm cm}^{-2}]^{-0.75}$,
for $N_{\rm H2} \geq 10^{14}$~\cd.  Combining $S_{\rm H2} = A N_{\rm H2}^{-0.75}$ 
($A = 3.16 \times 10^{10}~{\rm cm}^{3/2}$) with the approximation, 
$f \approx 2n_{\rm H2}/n_{\rm H}$ for $f \leq 0.1$, we can write the local molecular density as
\begin{equation} 
     n_{\rm H2}  =  \left( \frac {Rn_{\rm H}^2} {f_{\rm diss} \beta_0 G  A} \right) 
                     N_{\rm H2}^{0.75} \,e^{\tau_d}     \;  .
\end{equation}
In a planar slab of constant density $n_{\rm H}$, with column density $N_{\rm H2}$ at a distance 
$x$ into the absorber, we can write $n_{\rm H2} = dN_{\rm H2}/dx$.  The dust optical depth is
$\tau_d(x)  \approx n_{\rm H} x/N_d$, where $N_d \approx 4.5 \times10^{20}~{\rm cm}^{-2}$ 
at $\lambda \approx 1000-1100$~\AA\ (Draine 2011).  This leads to a differential equation,
\begin{equation}
 \frac {dN_{\rm H2}} {dx} = \left( \frac {Rn_{\rm H}^2} {f_{\rm diss} \beta_0 G  A} \right) 
           N_{\rm H2}^{0.75}  \, e^{n_{\rm H} x/N_d}  \; , 
\end{equation} 
with the analytic solution 
\begin{equation}
    N_{\rm H2} =  \left( \frac {R N_d n_{\rm H}} {4 f_{\rm diss} \beta_0 G A} \right)^4 
              \left[ e^{ \tau_d(x)} - 1 \right]^4 \; .
\end{equation}
The power-law approximation for $S_{\rm H2} \propto N_{\rm H2}^{-0.75}$ breaks down at low 
column densities, since $S_{\rm H2} = 1$ at $N_{\rm H2} < 10^{14}$~\cd.  However, this only occurs 
in a small surface layer ($f_{\rm H2} < 10^{-5}$).  Thus, the above expressions are valid up to to 
$f_{\rm H2} = 0.1$ for the \Htwo\ absorbers observed with \FUSE.  The observed atomic-to-molecular
transition at $N_{\rm H2} \approx 10^{19.5}~{\rm cm}^{-2}$ occurs at dust optical depth 
$\tau_d = N_{\rm H}/N_d$ given by 
 \begin{equation}
    \left[ e^{\tau_d} - 1 \right] = (1.755) \left( \frac {N_{\rm H2}} {10^{19.5}~{\rm cm}^{-2} } \right)^{1/4}  
          G \,  n_{30}^{-1}
         \left( \frac {3 \times 10^{-17}~{\rm cm}^3~{\rm s}^{-1} } {R} \right) 
        \left(  \frac {4.5 \times 10^{20}~{\rm cm}^{-2} } {N_d}  \right)  
 \end{equation}
Here, we adopted $f_{\rm diss} = 0.15$ and $A/N_d = 7.03 \times 10^{-11}~{\rm cm}^{1/2}$ for 
metallicities and grain opacities in the local ISM.   We scaled the cloud density to 
$n_{\rm H} = (30~{\rm cm}^{-3}) n_{30}$, appropriate for thermal pressures, 
$P/k \approx 3000~{\rm cm}^{-3}~{\rm K}$, inferred from observations of  \CI\ fine-structure 
populations (Jenkins \& Tripp 2011).   Setting the parenthetical terms equal to unity, we find a 
dust optical depth $\tau_d = 1.013$.  The right-hand side of equation~(8) scales as 
$(G/ R \, N_d \, n_{\rm H})$, which is insensitive to metallicity if the product ($R N_d$) remains 
constant.  This would be expected if grain abundances decrease at lower metal abundances.  
In that case, $N_d$ would increase and $R$ would decrease.  Thus, the transition should be 
governed by the FUV/density ratio ($G/n_{\rm H}$).   In low-metallicity environments such as the 
LMC/SMC, the transition will occur at similar $\tau_d \approx 1$, but at higher total hydrogen 
column density, $N_{\rm H} = \tau_d N_d$.

Previous studies of \Htwo\ abundances estimated that 
$R \approx 3 \times 10^{-17}~{\rm cm}^3~{\rm s}^{-1}$ (Jura 1975a).  The observed transition 
implies a rate coefficient,
\begin{equation}
   R  = \left[ \frac {4 f_{\rm diss} \beta_0 G A} {N_d n_{\rm H}} \right] N_{\rm H2} ^{1/4} 
                             \left[ e^{\tau_d} - 1 \right]^{-1} 
       \approx (3.0\times10^{-17}~{\rm cm}^3~{\rm s}^{-1}) G n_{30}^{-1}
                      \left[ \frac {N_{\rm H2}} {10^{19.5}} \right]^{1/4}    \; ,
\end{equation}
for $\tau_d \approx 1$, as found above.  This result suggests that absorbers with densities 
$n_{\rm H} > 30$~cm$^{-3}$ are associated with elevated radiation fields ($G > 1$) from their 
proximity to hot stars.  

We see that dust can be an important factor, along with \Htwo\ self-shielding, in the onset 
of the molecular transition.  Our results  show that the transition occurs when $\tau_d \approx 1$.  
Because the dust-to-gas ratio depends on abundances of C, Si, O, and other refractory heavy 
elements, changes in metallicity will have offsetting effects on the \Htwo\ fractions through the
formation coefficient ($R$) and radiative attenuation ($\tau_d$).  This analysis may explain the 
observed hydrogen column densities of the molecular transition at $f_{\rm H2} = 0.1$, which 
occurs at $\log N_{\rm H} \approx 21.0$ in the Milky Way,  $\log N_{\rm H} \approx 21.3$ in the LMC, 
and $\log N_{\rm H} \approx 22.0$ in the SMC.  Sight lines at high Galactic latitude 
(Gillmon \etal\ 2006) showed a transition at lower columns ($\log N_{\rm H} \approx 20.4$), 
likely because of lower radiation-to-density ratios, and possibly one-sided absorber illumination
from the disk stars.  The current \FUSE\ survey of the Milky Way disk exhibits no obvious shift 
with distance, but  large changes in metallicity are not expected over the range of target star 
distances.

More precise determinations of $R$ require knowledge of $n_{\rm H}$,  FUV radiation field, and 
dust properties, which may depend on metallicity and environment.  Interstellar clouds are 
inhomogeneous, with internal variations in temperature, hydrogen density, and metallicity.  
Irradiated cloud models have been constructed (Browning \etal\ 2003;  Le Petit \etal\ 2006;  
Nehm\'e \etal\ 2008; Klimenko \& Balashev 2020) that follow the attenuation of UV radiation 
into the cloud and the resulting changes in gas temperature ($T_g$), dust  temperature ($T_d$) 
and molecular fraction.  These models depend sensitively on $n_{\rm H}$, $\beta_0$, $G$, and 
$R(T_g, T_d, Z)$.  The best measures of the FUV radiation field and cloud density are the 
high-$J$ excitation ratios, $N(4)/N(2)$ and $N(5)/N(3)$.  These issues are discussed further 
in Section~3.5.   

\subsection{Rotational Excitation Temperatures ($J = 0,1,2$) }

The excitation temperature $T_{01}$ of the lowest two rotational states, $J = 0$ (para-\Htwo)
and $J = 1$ (ortho-\Htwo) is frequently used as a measure of the kinetic temperature in diffuse 
clouds.  This requires that the gas density and column density be sufficiently high for thermal 
proton collisions (Gerlich 1990) to couple the ortho and para forms of \Htwo\ and set the ratio 
$N(1)/N(0)$.   Recent experiments suggest that ortho-para conversion might occur on silicate 
grain surfaces (Tsuge \etal\ 2021).  The expectation is that ortho/para production on grains 
would be in the 3:1 spin-statistical ratio.
The observed low values of $T_{01} \approx 70-90$~K suggest that this formation channel
is subdominant. Theoretical models of \Htwo\ formation and destruction in diffuse clouds find
that the lowest three rotational states ($J = 0, 1, 2$) are usually thermalized, with populations 
depending primarily on gas temperature.   We assume that the observed populations obey 
Boltzmann ratios, with $T_{01}$ and $T_{02}$ determined from the expressions,
\begin{eqnarray}
          T_{01} &=& \frac { \Delta E_{01}/k} {\ln[(g_1/g_0)N(0)/N(1)]}   \; ,  \\
          T_{02} &=& \frac { \Delta E_{02}/k} {\ln[(g_2/g_0)N(0)/N(2)]}   \; .  
\end{eqnarray}
Here, $g_1/g_0  = 9$ and $g_2/g_0 = 5$ are ratios of statistical weights of the rotational levels,
and $\Delta E_{01}/k = 170.48$~K and $\Delta E_{02}/k = 509.86$~K come from the rotational 
energies computed by Komasa \etal\ (2011).  

The initial {\it Copernicus} study of \Htwo\ in 13 clouds with $N(0) \geq 10^{17}~{\rm cm}^{-2}$ 
found a mean temperature $\langle T_{01} \rangle = 81 \pm 13$~K (Spitzer \& Cochran 1973).  
A more extensive {\it Copernicus} survey (Savage \etal\ 1977) of 61 stars with 
$\log N_{\rm H2} \geq 18$ found $\langle T_{01} \rangle = 77 \pm 17$~K (rms).   
In their study of CO and \Htwo\ in diffuse molecular clouds, Sheffer \etal\ (2008) tabulated  
total \Htwo\ column densities $N_{\rm H2}$ along 58 sight lines studied with \FUSE.  For 24 
stars in common with our 139, there was reasonable agreement in $\log N_{\rm H2}$, typically 
within $\pm0.10$.  We also compared our \Htwo\ column densities to those in the metallicity
study (O, Ge, Kr) of Jenkins (2019). For 57 stars in common, we found near agreement
(within $\pm 0.03$) for $\log N(0)$ and $\log N(1)$ in 22 and 18 sight lines, respectively.  For 
the other stars, the column densities agreed within $\pm 0.10$.  We also compared the tabulated
rotational temperatures in these two studies.  Sheffer \etal\ (2008) quoted a mean rotational 
temperature, $\langle T_{01} \rangle = 76 \pm 14$~K for 56 sight lines, similar to the 
{\it Copernicus} result  of  $77\pm 17$~K (Savage \etal\ 1977).  In our current \FUSE\ survey 
of 139 stars, we find $\langle T_{01} \rangle = 88 \pm 20$~K.  Comparing to the 57 common 
sight lines with Jenkins (2019), we found only a few differences.  These are attributable to different 
values of $N(0)$ and $N(1)$ arising from fitting the damping wings.   Sheffer \etal\ (2008) did 
not tabulate individual column densities, $N(0)$ and $N(1)$, and we were unable to investigate
the source of the differences. 

{\bf Figure~7}  shows the distributions of $T_{01}$ with $E(B-V)$ and $N_{\rm H}$.  Removing 
the four labeled outliers reduces the mean slightly to $\langle T_{01} \rangle = 87$~K.   The 
distributions of $T_{02}$ are shown in {\bf Figure~8}.  For the reduced sample (four outliers 
removed) of 128 stars with measured $J = 2$ column densities, we find 
$\langle T_{02} \rangle = 77 \pm 18$~K.   
In their analysis of \Htwo\ in the LMC and SMC,  Tumlinson \etal\ (2002) found 
$\langle T_{01} \rangle = 82 \pm 21$~K for 22 sight lines with $N_{\rm H2} \geq 10^{16.5}$~\cd.  
Kruczek \etal\ (2019) explored the contributions of higher rotational lines to the damping wings
of $J = 1$ lines.  Their re-analysis of nine {\it Copernicus} sight lines and 13 from \FUSE\ altered 
$N(0)$ and $N(1)$, resulting in a reduced $T_{01} = 68\pm13$~K (12\% lower than their previous 
values).  Our \FUSE\ survey extends to larger column densities and greater stellar distances than 
{\it Copernicus} and contains more than twice the number of stars.  Given the dispersions in 
$T_{01}$ and $T_{02}$ distributions, their agreement suggests that the diffuse ISM has similar 
heating and cooling rates over a wide range of cloud densities, metallicities, and FUV radiation fields.  

Even for the best determinations of $\log N(0)$ and $\log N(1$), rotational temperatures $T_{01}$ 
have errors of $\sigma_{T_{01}} \approx 4$~K.  Uncertainties on $T_{02}$ can be higher,  
when $\log N(2)$ is poorly determined.    From errors on $\log N_0$, $\log N_1$, and $\log N_2$, 
and neglecting covariance, the propagated errors on $T_{01}$ and $T_{02}$ derived from 
equations (10) and (11) are:
\begin{eqnarray} 
    \frac {\sigma_{T_{01} } } { T_{01}} &=&  2.303  \left[ \frac { T_{01} } {170.48~{\rm K} } \right] 
               \left[ \sigma_{\log N_0}^2 + \sigma_{\log N_1}^2 \right]^{1/2}    \\
 \frac {\sigma_{T_{02} } } { T_{02}} &=&  2.303  \left[ \frac { T_{02} } {509.86~{\rm K} } \right]  
               \left[ \sigma_{\log N_0}^2 + \sigma_{\log N_2}^2 \right]^{1/2}  \;  .               
\end{eqnarray} 
For equal errors on $J = 0$ and $J = 1$ column densities, $\sigma_{\log N_1} =  \sigma_{\log N_0}$,
this expression simplifies to $\sigma_{T_{01}} \approx [T_{01}^2 / 52.34] \sigma_{\log N_0}$.  For the 
best fits to the damping wings ($\sigma_{\log N_0} = 0.03$) the temperature uncertainty is
$\sigma_{T_{01}} = 4.3$~K at $T_{01} \approx 87$~K.  In poorer quality data, with
$\sigma_{\log N_0}  \approx \sigma_{\log N_1} = 0.07$, the uncertainty is higher, 
$\sigma_{T_{01}} \approx 10$~K.  Errors on $J = 2$ column densities are usually much larger 
than those for $J = 0$.  Thus, $\sigma_{T_{02}} \approx [T_{02}^2 / 221.39] \sigma_{\log N_2}$.  
At $T_{02} \approx 77$~K, the uncertainty $\sigma_{T_{02}} \approx$ 5--11~K for 
$\sigma_{\log N_2} = 0.20-0.40$.

\subsection{Rotational Excitation Temperatures ($J \geq 3$) }

The higher rotational levels ($J \geq 3$) of \Htwo\ are generally believed to be populated by the
 fluorescent cascade following FUV radiative pumping in the Lyman and Werner bands
(Black \& Dalgarno 1976;  Spitzer \& Zweibel 1974; Jura 1974).  The FUV radiation field includes 
the ambient Galactic radiation field (Jura 1974; Habing 1968; Draine 2011) augmented by local 
flux from O stars.   Measurements by {\it Copernicus} (Spitzer \etal\ 1974; Morton 1975) found that 
levels $J \geq 3$ were populated in excess of predictions from $T_{01}$.  The column densities in 
$J = 3, 4, 5$ were fitted to excitation temperatures $T_{\rm exc} \approx 200-500$~K, and sometimes
as high as 1100~K near luminous early O-type stars such as Zeta Puppis (Morton \& Dinerstein 1976).
Subsequent  studies of  \Htwo\ excitation using \FUSE\ data (Browning \etal\ 2003; Sheffer \etal\ 2008; 
Nehm\'e \etal\  2008; Jensen \etal\ 2010) also found $T_{\rm exc} \approx 200-500$~K for $J \geq 3$.   
The mean rotational excitation temperature, fitted to $J \geq 3$ and averaged over our sample, 
is  $T_{\rm exc} = 326 \pm 125$~K with a typical range from 150--650~K.  These temperatures are 
similar to those seen toward selected {\it Copernicus} targets, and they likely reflect fluorescent  
pumping of high-$J$ states.  Some observations suggest that $J \geq 2$ levels may be influenced 
by collisional excitation (Gry \etal\ 2002; Nehm\'e \etal\ 2008; Ingalls \etal\ 2011) in a component of 
warm gas  ($T > 500$~K) heated by turbulent dissipation (Moseley \etal\ 2020).  Observationally,
CH$^+$ is correlated with rotationally-excited \Htwo\ (Jensen \etal\ 2010), leading to suggested
production schemes for CH$^+$ involving  hot \Htwo\ (Falgarone \& Puget 1995; Myers \etal\ 2015).
Sheffer \etal\ (2008) tabulated several excitation temperatures to higher rotational states ($T_{03}$
and $T_{04}$) for 56 \FUSE\ sight lines.  We chose not to tabulate these parameters, since our 
modeling experience (Browning \etal\ 2003) found them less useful than $T_{24}$ and $T_{35}$. 
Many of their values of $T_{04}$ seemed implausibly high (200-300~K) and inconsistent with our 
observed  ratios of $N(4)/N(0)$. Because they did not tabulate individual values of $N(J)$, we 
could not investigate further.

In our survey, we choose to focus on individual pairs of upper ($u$) and lower ($l$) rotational states 
$(J_l, J_u)$, in particular (0,2), (2,4), and (3,5).  These parameters capture the fact that \Htwo\ 
exists over a range of cloud temperatures, with radiative pumping changing throughout the 
cloud because of \Htwo\ self-shielding and attenuation by  FUV extinction.   In addition, the rate of 
ortho-para conversion may change, depending on cloud density.  This will affect radiative pumping
from $J = 0$, $J = 1$, and sometimes $J = 2$, creating departures of rotational populations of
$J \geq 3$ from a single excitation temperature.  For these reasons, we tabulate pairwise excitation
temperatures, $T_{02}$, $T_{24}$, and $T_{35}$, each remaining within para (even-$J$) and ortho
(odd-$J$) forms of \Htwo.  The temperatures $T_{04}$ and $T_{15}$ are not as useful diagnostics.  
Detailed models of the excitation processes and radiative transfer may help to distinguish the 
relative contributions of FUV pumping and collisional excitation and to estimate the FUV 
radiation field and gas density.  
 
Based on column densities $N(J)$ in the \FUSE\ survey, {\bf Table~4} lists four excitation temperatures, 
$T_{01}$, $T_{02}$, $T_{24}$, and $T_{35}$, between upper and lower rotational states defined by
\begin{equation}
   T_{lu} = \frac { (E_u - E_l) /k }  { \ln [( g_u /g_l) (N_l/N_u)] }   \; , 
\end{equation} 
corresponding to Boltzmann population ratios, 
\begin{eqnarray}
  N(1)/ N(0) &=&  (9/1) \exp [ -170.48~{\rm K} / T_{01} ]    \\   
  N(2)/ N(0) &=&  (5/1) \exp [ -509.86~{\rm K} / T_{02} ]    \\
  N(4)/ N(2) &=& (9/5)  \exp [ -1171.78~{\rm K} / T_{24} ]   \\
  N(5)/ N(3) &=& (11/7)  \exp [ -1488.66~{\rm K} / T_{35} ]   \;  .  
\end{eqnarray} 
These excitation temperatures were derived from the relativistic quantum calculations of 
Komasa \etal\ (2011), using the $J$-level dissociation energies in their Table~1.  From the 
observed populations of higher-$J$ states, we find mean excitation temperatures, 
$\langle T_{24} \rangle = 237 \pm 91$~K  and $\langle T_{35} \rangle =  304 \pm 108$~K.   
Even with a wide range of these temperatures, they generally exhibit a correlation between 
the two parameters.  

{\bf Figure 9} shows the rotational distributions in four of the six sight lines with detections 
up to $J = 7$.  These include Star \#48 (HD~93250) with $\log N(7) = 15.54 \pm 0.15$;  
Star \#88 (HD~163892) with $\log N(7)= 14.14 \pm 0.06$;
Star \#116 (HD~199579) with $\log N(7)= 14.29 \pm 0.11$; and 
Star \#137  (HD~303308) with $\log N(7)= 15.43 \pm 0.15$.   The absorbers along these
sight lines have different excitation temperatures.  Two exhibit higher excitation
temperatures for $J = 4$ and $J = 5$, likely produced by elevated FUV radiation fields 
from absorber proximity to hot stars.

{\bf Figure 10} displays distributions of the four excitation temperatures with stellar target 
distance.  We see no obvious trend with increasing distance, although stars beyond 
4--5~kpc are unlikely to be an unbiased sample.  The similar values of  of $T_{01}$ and 
$T_{02}$ suggest that  the $J = 0, 1, 2$ levels are thermalized to the gas kinetic temperature.  
Both $T_{24}$ and $T_{35}$ show several outliers, well above their distribution means 
of 237~K and 304~K, respectively.  For this 10-15\% population, higher rotational excitation 
is expected from exposure to local FUV radiation above the background.

 {\bf Figure 11} shows the relation of $T_{24}$ and $T_{35}$, color-coded by the SpT of the
 target star.  With the exception of a few labeled outliers, the para-\Htwo\ levels 
 ($J = 2$ and 4) and ortho-\Htwo\ levels ($J = 3$ and 5) have correlated excitation
 temperatures above 300~K.   Most sight lines have $T_{35} > T_{24}$, with data points 
 above the dashed line of unit slope.  This difference could arise from strong pumping of 
 ortho-\Htwo\ populations out of $J = 1$. Alternatively, thermalization of the $J = 2$ population
 to the gas kinetic temperature may alter the pumping out of $J = 2$.  The expectation that 
 high-$J$ populations would be greater toward hotter O-type stars is not consistently reflected 
 in this plot.  The labeled sight lines with $T_{35} > 500$~K include two early O-type stars 
 (\#4, \#48), two later O-type stars (\#32, \#122), and two B0/B0.5 stars (\#2, \#59).  A number
 of hot (O2--O4) stars exhibit high excitation temperatures ($T_{35} > 450$~K), while others 
 have $T_{35} < 400$~K.  This mixed distribution suggests that radiative pumping of $J \geq 3$ 
 levels is sometimes enhanced by proximity of the molecular absorbers to nearby hot stars.  
 Deriving the pertinent model parameters of FUV absorption rate and absorber density will 
 require modeling of the ratios $N(4)/N(2)$ and $N(5)/N(3)$, as performed by Browning \etal\ 
 (2003) and Klimenko \& Balashev (2020).




\section{Discussion and Summary}

This \FUSE\ survey of interstellar \Htwo\ in the Milky Way disk complements the pioneering 
survey by {\it Copernicus}, with several important extensions.   Because \FUSE\ was a more 
sensitive spectrograph,  the survey includes more OB-star targets at greater distances and 
larger \Htwo\ column densities.  Many \FUSE\ sight lines exhibit molecular fractions above the 
value, $f_{\rm H2} \geq 0.01$ at  $\log N_{\rm H} \geq 20.7$, noted in the {\it Copernicus} survey 
(Savage \etal\ 1977).   We also measured \Htwo\ populations in higher rotational states
($J \geq 2$) as well as $J = 0$ and $J = 1$.  The \FUSE\ sample includes the 139 OB-star 
targets with updated distances and $E(B-V)$ from Shull \& Danforth (2019) derived from 
updated spectral types, digital photometry, and optical-NIR dust extinction in the Galactic 
O-star Spectroscopic Survey (Ma\'iz Apell\'aniz \etal\ 2004; Ma\'iz-Apell\'aniz \& Barb\'a 2018).   
Our measurements of \Htwo\ populations in the lowest three rotational states ($J  = 0, 1, 2$) 
found similar excitation temperatures, $T_{01} \approx T_{02} \approx 70-90$~K,  
suggesting thermal coupling to the gas kinetic temperature.  Populations of higher rotational 
states ($J \geq 3$) could be use to distinguish between radiative pumping and collisional
excitation.  The pairwise excitation temperatures, $T_{24}$ and $T_{35}$, are correlated
at the high end of the distribution, with $T_{35} > T_{24}$.  After deriving total hydrogen 
column densities, $N_{\rm H}$, from those of \Htwo\ and \HI, we compared them to updated 
values of selective extinction, $E(B-V)$, to find a mean gas-to-dust ratio in the Galactic disk, 
$\langle N_{\rm H}/E(B-V) \rangle = (6.07 \pm 1.01) \times 10^{21}~{\rm cm}^{-2}~{\rm mag}^{-1}$.  

The \FUSE\ survey finds that a typical atomic-to-molecular transition in the ISM of the Galactic 
disk occurs at molecular fraction $f_{\rm H2} \approx 0.1$ and molecular column density 
$N_{\rm H2} \approx 10^{19.5}~{\rm cm}^{-2}$.  This transition can be understood with a simple 
analytic model describing its dependence on \Htwo\ formation on dust-grain surfaces and 
photo-dissociation by FUV radiation, including self-shielding and dust attenuation.  This 
formulation shows that the transition depends on the ratio of FUV flux to gas density, 
analogous to the photoionization parameter used in the analysis of nebular lines.
The transition occurs at dust optical depth $\tau_d \approx 1$ and total hydrogen column 
density $N_{\rm H} = \tau_d N_d$, where the dust-opacity parameter 
$N_d \approx 4.5 \times 10^{20}~{\rm cm}^{-2}$ at solar metallicity ($Z \approx Z_{\odot}$).    
The \HI-to-\Htwo\ conversion is mediated by both dust opacity and \Htwo\ self-shielding. 
The optical depth $\tau_d \approx 1$ is likely insensitive to metallicity and grain/gas ratio,
because of offsetting effects of grain surface area on the \Htwo\ formation rate coefficient 
($R$) and dust opacity ($N_d$).   

With a mean fractional abundance $\langle f_{\rm H2} \rangle \approx 0.2$ in diffuse interstellar 
clouds, radiative cooling by 28.22~$\mu$m emission from the $J = 2$ level of \Htwo\ (510~K 
excitation) augments the dominant cooling from the 157.74~$\mu$m [\CII] fine-structure line 
(91.21~K excitation).  At low densities, the cooling rate from H$^{\circ}$-\Htwo\ collisions is 
${\cal L}_{\rm H2} = (3.1 \times 10^{-28}$ erg~cm$^3$~s$^{-1}) n_{\rm HI}  n_{\rm H2}$
at $T =100$~K (Forrey \etal\ 1997), while that from [\CII] is 
${\cal L}_{\rm CII} \approx (3.0 \times 10^{-28}$ erg~cm$^3$~s$^{-1}) n_{\rm H}^2$.
We adopted an electron-impact collision strength $\Omega_{12} = 1.56$, a solar carbon 
abundance $n_C  /n_{\rm H} = 2.69 \times 10^{-4}$, and electron density 
$n_e \approx 3.3 \times 10^{-4} n_{\rm H}$ donated by trace metal ions.  
Thus, ${\cal L}_{\rm H2} / {\cal L}_{\rm CII} \approx 0.5 f_{\rm H2} (1-f_{\rm H2})$ at 100~K.  
In higher density clouds, the \Htwo\ 28.22~$\mu$m emission will be reduced by
collisional de-excitation.   The \Htwo\ cooling rises at $T > 100$~K, but drops off 
exponentially in lower temperature clouds at higher $N_{\rm H}$, owing to the 510~K 
excitation temperature of the $J = 2$ level.   

\noindent
The following summarizes the results of the \FUSE\ survey of interstellar \Htwo\ abundances 
and inferred physical parameters in the Milky Way disk:
\begin{enumerate}

\item The \FUSE\ survey measured column densities of \Htwo\  in the Galactic disk toward 
139~OB stars with recently updated SpT, photometry, and distances (Shull \& Danforth 2019).
The survey extends the {\it Copernicus} \Htwo\ survey (Savage \etal\ 1977) up to total hydrogen 
column densities $N_{\rm H} \approx 5 \times 10^{21}$ cm$^{-2}$ and complements  \FUSE\ 
surveys of \Htwo\ in the Magellanic Clouds (Tumlinson \etal\ 2002), translucent clouds 
(Rachford \etal\ 2002, 2009), and gas at high Galactic latitude (Gillmon \etal\ 2006; Wakker 2006).  

\item For each sight line, we report column densities $N_{\rm H2}$, $N_{\rm HI}$, $N(J)$, 
$N_{\rm H} = N_{\rm HI} + 2N_{\rm H2}$, and $f_{\rm H2} = 2N_{\rm H2} / N_{\rm H}$,
with mean values listed in Table~5.  The mean gas-to-dust ratio,
$\langle N_{\rm H}/E(B-V) \rangle = (6.07\pm1.01) \times 10^{21}$ cm$^{-2}$~mag$^{-1}$, 
is slightly above the value of $5.8 \times 10^{21}$ cm$^{-2}$~mag$^{-1}$ in the {\it Copernicus} 
survey (Bohlin \etal\ 1978).  The larger ratios seen in 21-cm/far-IR surveys at high Galactic latitudes 
(Liszt 2014a,b) suggest different distributions of gas and dust above the disk, produced by 
grain sedimentation to the disk plane or dust destruction when transported above the disk.

\item Using an analytic model of \Htwo\ formation-destruction equilibrium with dust opacity 
and \Htwo\ self-shielding, we derive an expression for the atomic-to-molecular transition, 
which occurs at optical depth $\tau_d \approx 1$,  molecular fraction $f_{\rm H2} \approx 0.1$, 
$N_{\rm H2} \approx 10^{19.5}$~\cd, and $N_{\rm H} \approx 10^{21}$~\cd.   An \Htwo\ formation 
rate coefficient $R \approx 3 \times10^{-17}~{\rm cm}^3~{\rm s}^{-1}$ is consistent with the 
observed transition, with occasional elevated FUV radiation fields in 10-15\% of the sight lines.  
These parameters can be constrained with models of \Htwo\ rotational populations, supplemented 
by \CI\  fine-structure abundances (Jenkins \& Tripp 2011; Klimenko \&  Balashev 2020).

\item The lowest three rotational states ($J = 0, 1, 2$) appear to be thermally coupled by 
collisions to the gas kinetic temperature.  The survey mean excitation temperatures are
$\langle T_{01} \rangle = 88 \pm 20$~K and $\langle T_{02} \rangle = 77 \pm 18$~K.
For sight lines with $E(B-V) > 0.5$ and $\log N_{\rm H} > 20.7$, these temperatures 
decrease to $50-70$~K.  

\item Populations of higher-$J$ states are produced primarily by radiative pumping from 
FUV radiation.  From column-density ratios of rotational levels $N(4)/N(2)$ and $N(5)/N(3)$, 
we find mean excitation temperatures, $\langle T_{24} \rangle =  237 \pm 91$~K  and 
$\langle T_{35} \rangle = 304 \pm 108$~K (rms).  In most cases, these two temperatures
are correlated, with $T_{35} > T_{24}$, but  we find no consistent connection with SpT (from
O3 to B1).  Elevated radiative pumping is likely produced by close proximity to hot stars,
possibly with some collisional excitation from heating by turbulent dissipation.  

\end{enumerate}

\noindent
{\bf Acknowledgements.}  This work was supported by the \FUSE\ mission, with financial support 
from NASA Contract NAS5-32985 to Johns Hopkins University and a sub-contract to the 
University of Colorado at Boulder.  We thank Jason Tumlinson for developing software for 
\FUSE\ studies of \Htwo\ and former CU undergraduates Teresa Ross and Kristen Gillmon
for their assistance with data analysis during early stages of this project.  We have benefitted 
from discussions on interstellar gas, molecules, and dust with Sergei Balashev, John Black, 
Bruce Draine,  Kevin France, Ed Jenkins,  Slava Klimenko, Harvey Liszt,  Chris McKee,
Blair Savage, and Don York.




\clearpage


\begin{figure}
\includegraphics[angle=0,scale=0.75]{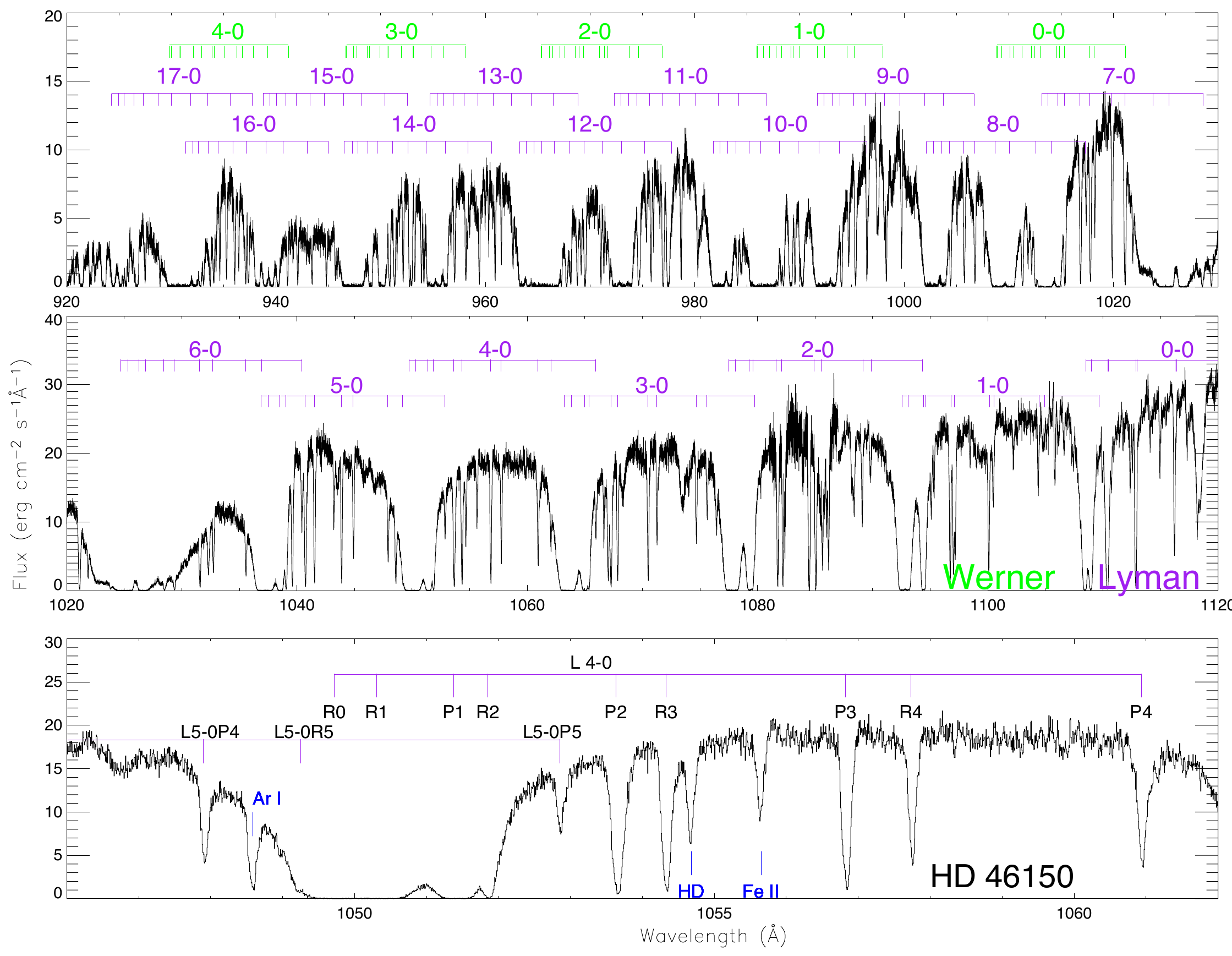}
\caption{\FUSE\ spectrum of the sight line to HD 46150, an O5~Vf star at  1.5~kpc distance and 
$E(B-V) = 0.45$, located inside the Rosette Nebula.   Locations of the H$_2$ Lyman and Werner 
bands are shown as purple and green templates in the top two panels. Each template shows a
different vibrational band ($v_u-v_{\ell}$) with a comb of absorption lines from rotational levels 
$J$ in the ground vibrational state ($v_{\ell} = 0$) to the upper vibrational state ($v_u$) in the 
excited electronic states, $B\,^{1}\Sigma_u^+$ (Lyman bands) and $C\,^{1}\Pi_u$ (Werner bands).  
For each band, the template shows locations of R-branch and P-branch absorption lines from
$J = 0, 1, 2,...$.  The bottom panel shows a close-up of the (4-0) Lyman band, with nearby lines of 
\ArI\ (1048.220~\AA), \FeII\ (1055.262~\AA), HD (4-0) R(0) at 1054.294~\AA, and  P(4), R(5), P(5) 
lines from $J = 4$ and $J = 5$ levels of the (5-0) Lyman band.  The R(0), R(1), P(1) lines have 
damping wings and are blended.  In some bands, the P(1) and R(2) lines are separable.  The
P-branch and R-branch lines from higher-$J$ states are sufficiently shifted to be measured.
We find $\log N_{\rm H2} = 20.64\pm0.04$,  with $\log N(0) = 20.23 \pm 0.05$, 
$\log N(1) = 20.42 \pm 0.05$, and rotational temperatures $T_{01} = 97 \pm 9$~K and 
$T_{02} = 55\pm4$~K.    }
\end{figure}

\clearpage


\begin{figure}
\includegraphics[angle=0,scale=0.6]{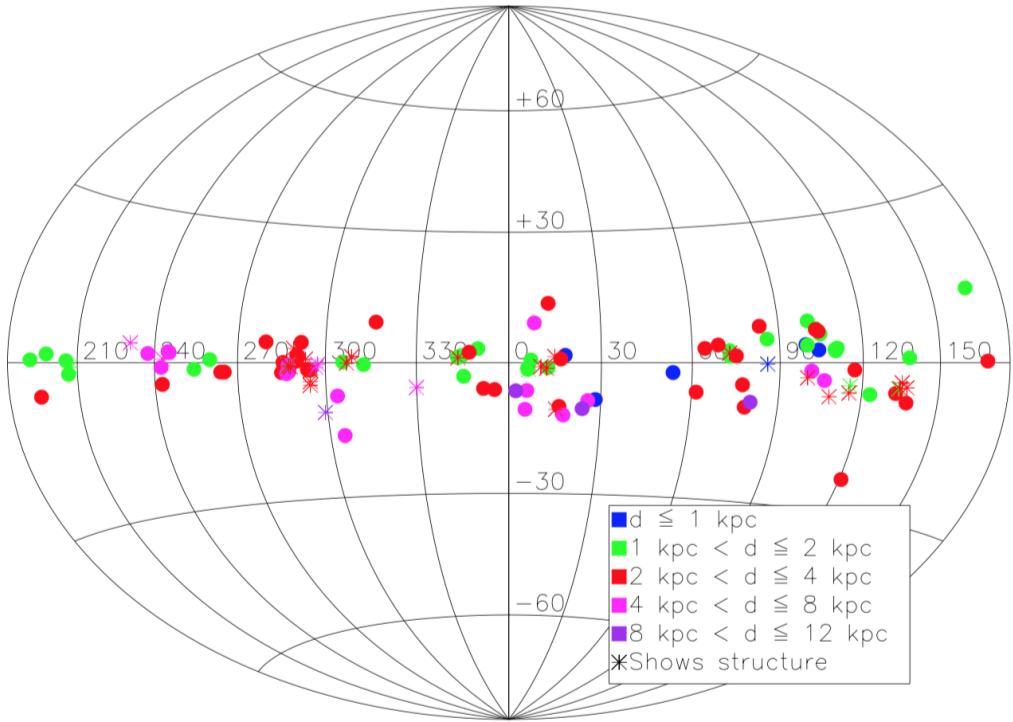}
\caption{Aitoff projection plot of 139 target locations in Galactic coordinates
($\ell$ and $b$) with distances coded by color.  Several sight lines with multiple 
velocity components that are separably measurable ($\Delta v \geq 20$~\kms) are 
shown as asterisks. }
\end{figure}

\begin{figure}[!htb]
\includegraphics[angle=0,scale=1.0]{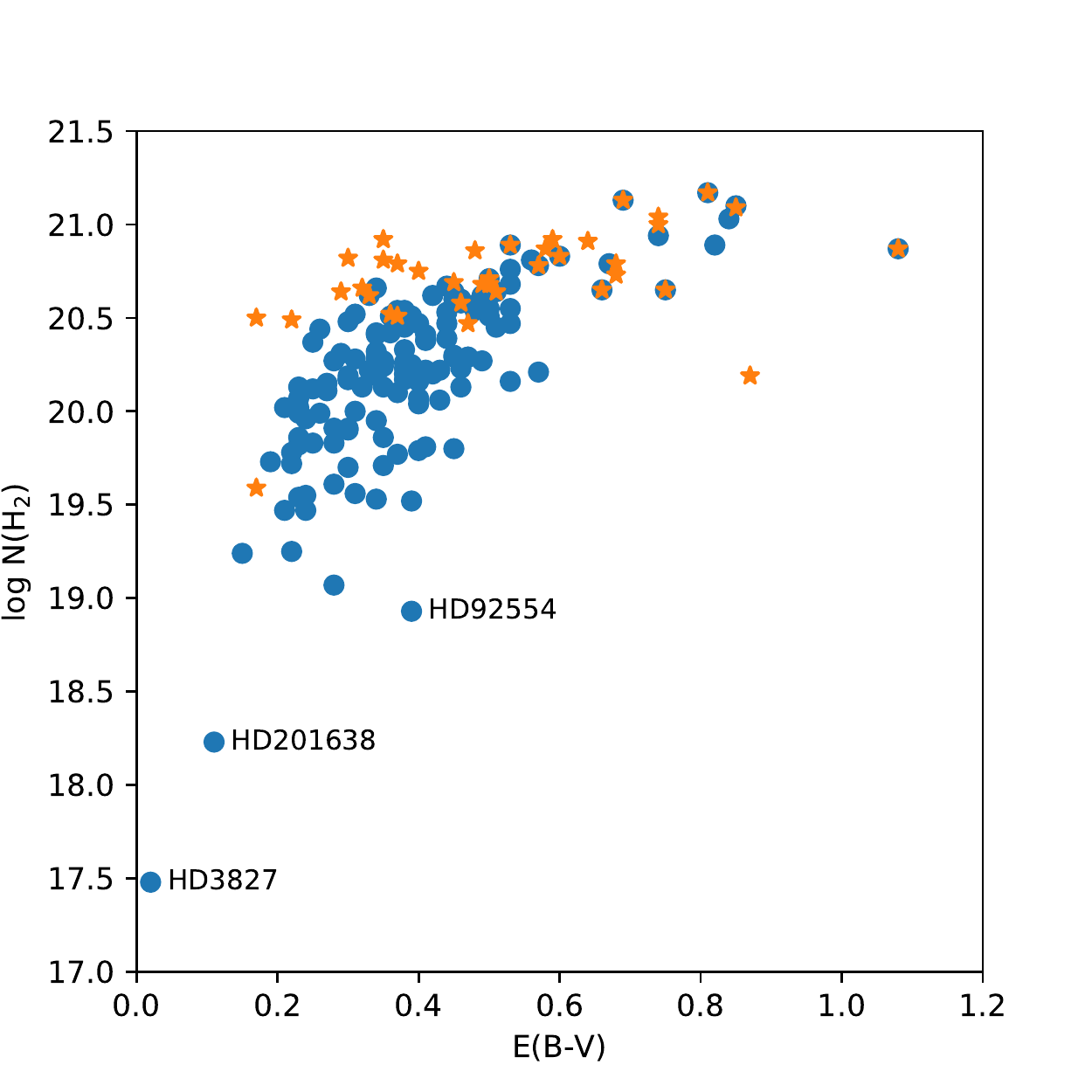}
\caption{Distribution of \Htwo\ column density with color excess, labeling three sight lines 
with $\log N_{\rm H2} < 19.0$.  Symbols are color-labeled for the current \FUSE\ 
survey (blue circles) and translucent sight lines (orange stars) also studied by \FUSE\
(Rachford \etal\ 2002, 2009).  We re-analyzed 11 of these translucent sight lines with
our \Htwo\ software for stars with new GOS photometry and SpTs
(ID numbers 32, 82, 97, 98, 105, 107, 114, 116, 120, 122, 127).   }  
\end{figure}

\clearpage


\begin{figure*}[!ht]
\includegraphics[angle=0,scale=0.55]{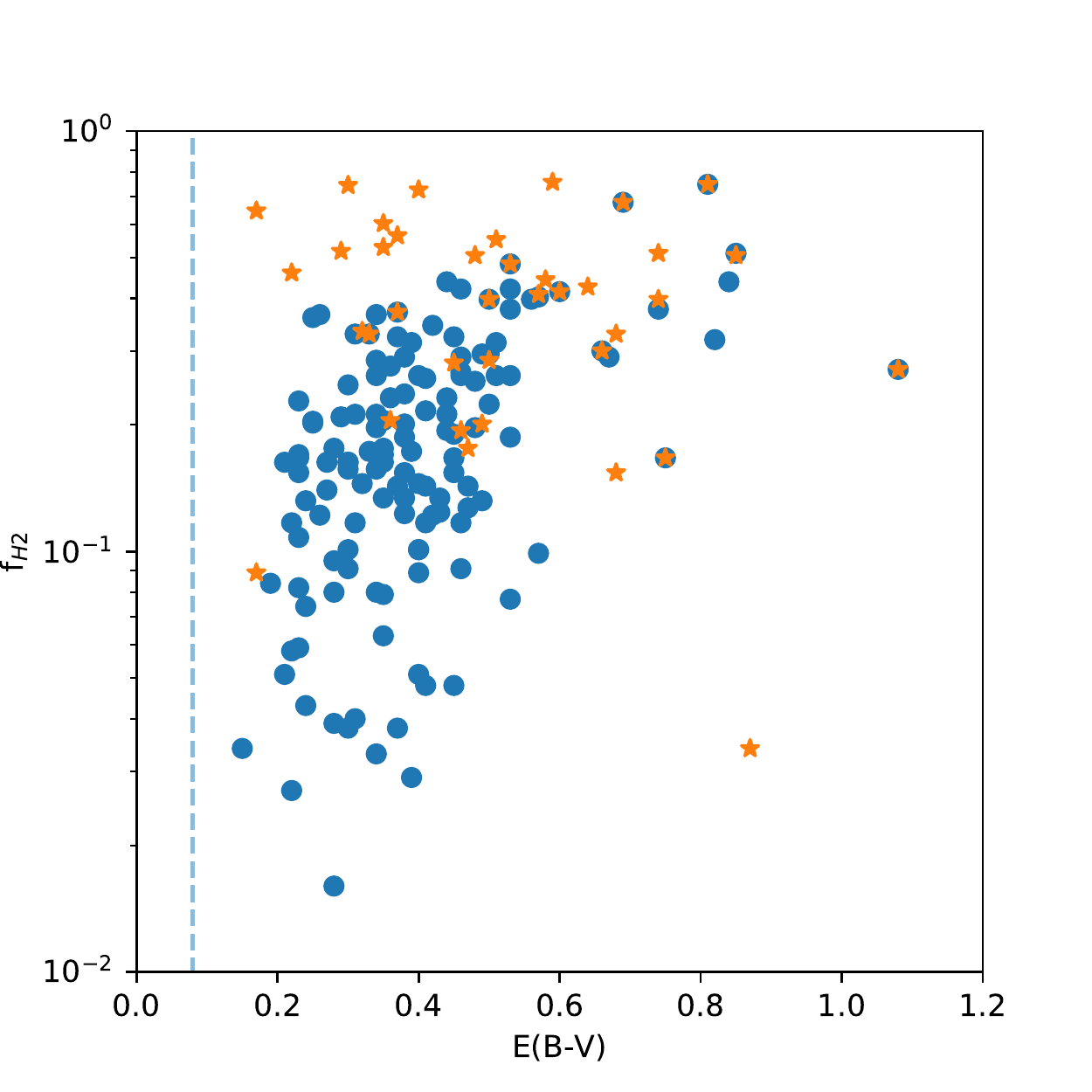}
\includegraphics[angle=0,scale=0.55]{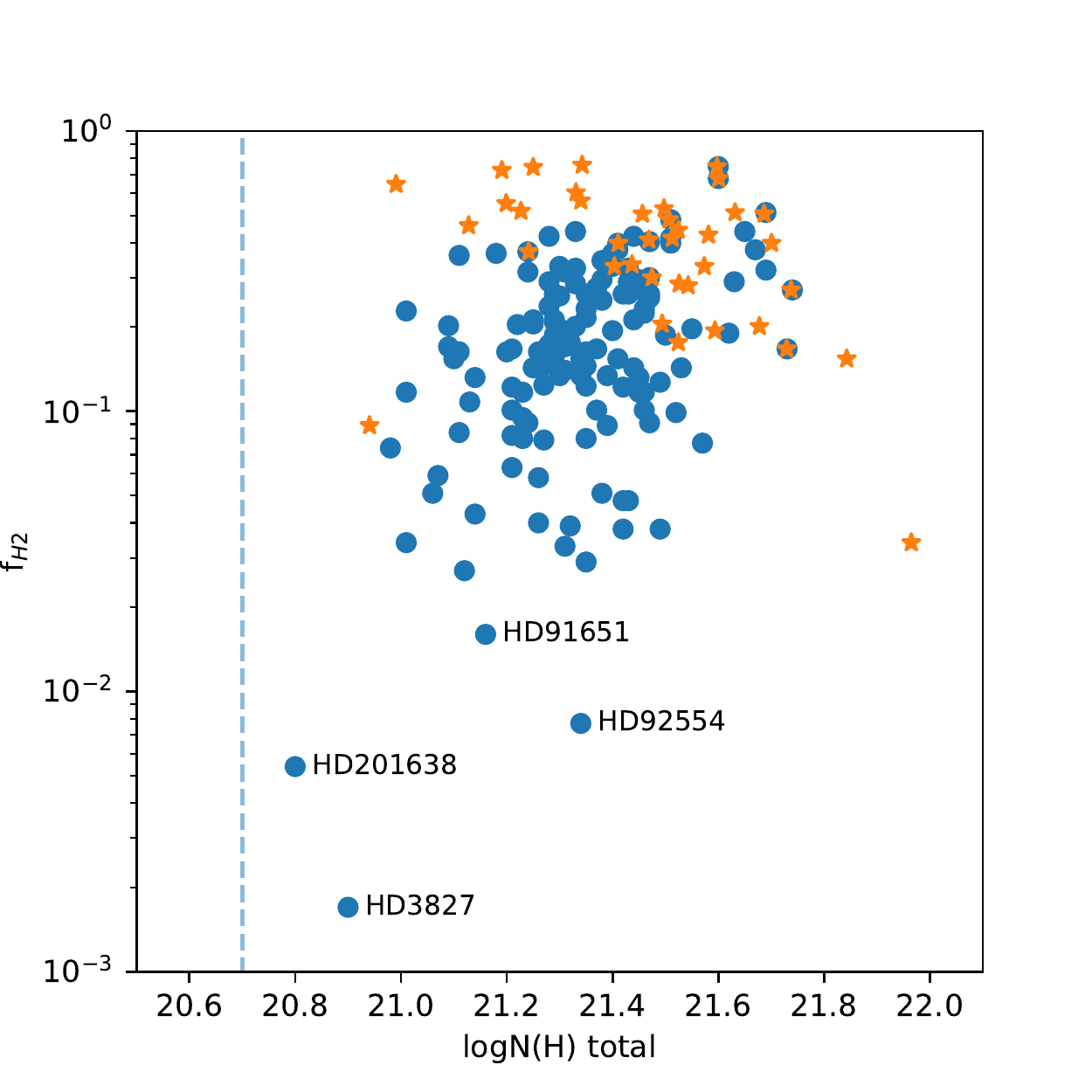}
\caption{Molecular fraction $f_{\rm H2}$ compared to color excess (left panel) and 
total hydrogen column density, $N_{\rm H} = N_{\rm HI} + 2 N_{\rm H2}$ (right panel).  
Vertical dashed lines show the transition to $f_{\rm H2} > 0.01$ at $E(B-V) \gtrsim 0.08$ 
and $\log N_{\rm H} \gtrsim 20.7$ seen in {\it Copernicus} data (Savage \etal\ 1977).  
Most of the \FUSE\ targets are more distant and have $f_{\rm H2}$ between 3\% and 75\%.  
Symbols are color-coded as in Figure 3.   The outlier (orange star) with $f_{\rm H2} = 0.034$ 
at $E(B-V) = 0.87$ and $\log N_{\rm H} = 21.96$ is the translucent sight line toward 
HD~164740 (Rachford \etal\ 2009).} 
\end{figure*}


\begin{figure*}[!hb]
\includegraphics[angle=0,scale=0.53]{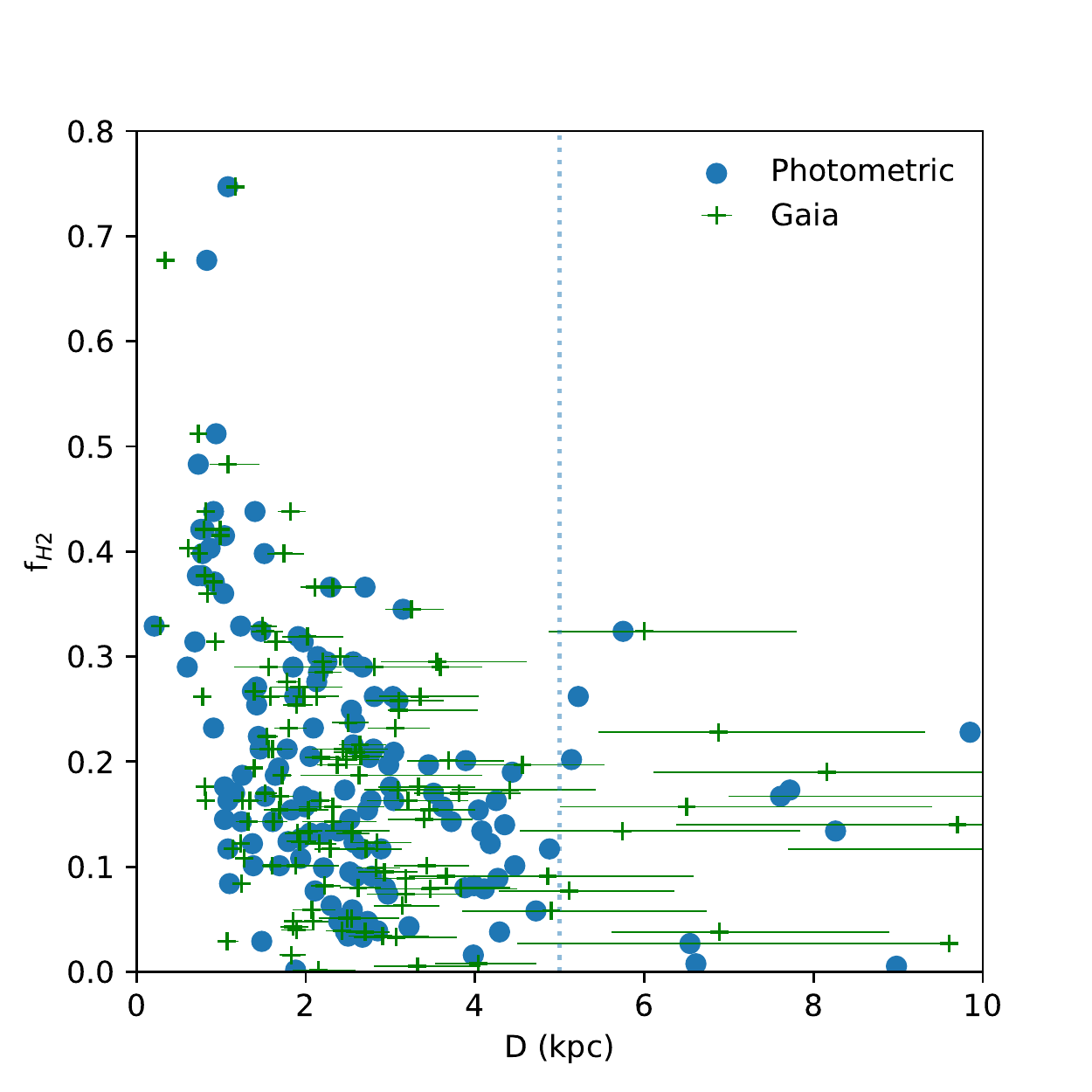}
\includegraphics[angle=0,scale=0.53]{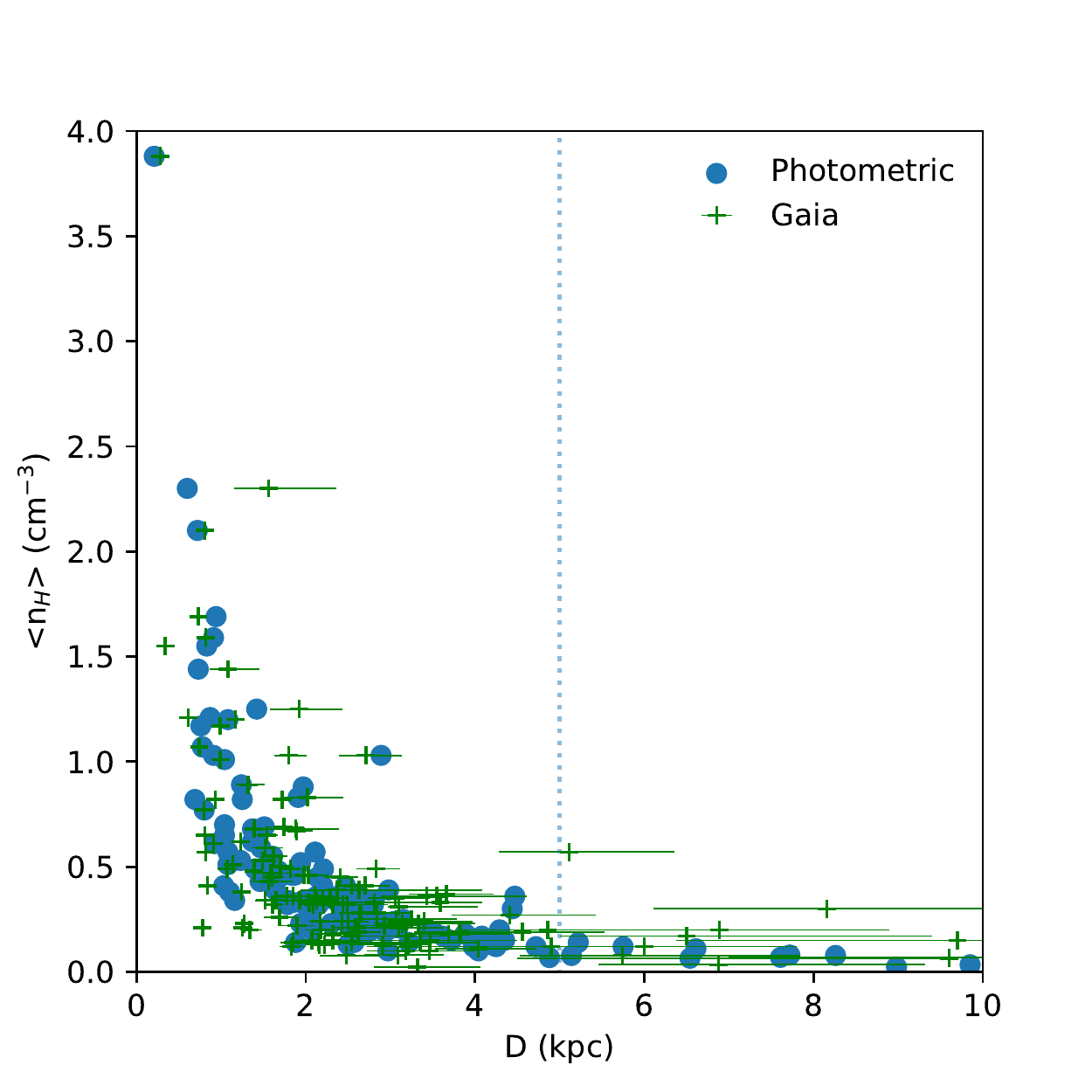}
\caption{Molecular fraction $f_{\rm H2}$ (left panel) and total hydrogen density $n_{\rm H}$ 
(right panel), averaged over photometric distance to target stars.  Updated values 
(Shull \& Danforth 2019) are shown for both photometric and {\it Gaia}-DR2 
parallax distances.   For 129 stars with photometric distances $D \leq 5$~kpc 
(vertical dashed lines)  the mean sight-line values are 
$\langle n_{\rm HI} \rangle = 0.50$~cm$^{-3}$, $\langle f_{\rm H2} \rangle = 0.20$, and
$\langle N_{\rm H} / E(B-V) \rangle = 6.07 \times 10^{21}~{\rm cm}^{-2}~{\rm mag}^{-1}$. 
Table 5 lists these quantities for the full survey and for sub-samples ($D \leq 2$~kpc 
and $D \leq 5$~kpc). }
\end{figure*}

\clearpage

\begin{figure}[!htb]
\includegraphics[angle=0,scale=1.2]{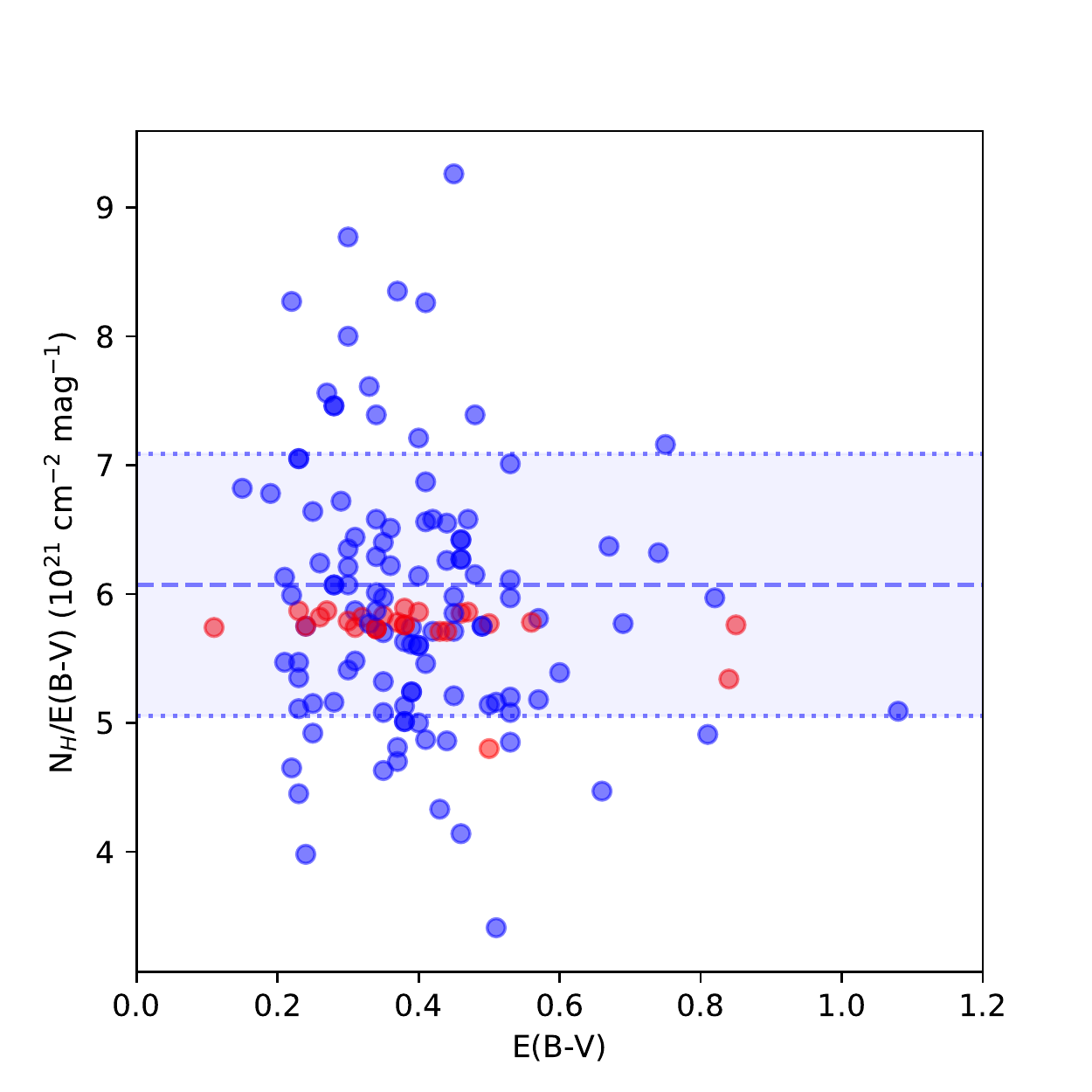}
\caption{Distribution of the ``gas-to-dust" ratio, $N_H/E(B-V)$, for 138 sight lines, in units
of $10^{21}$ cm$^{-2}~{\rm mag}^{-1}$.   One star (ID \#7, HD 3827) was omitted owing to 
its uncertain $E(B-V)$.
Blue points are the 112 sight lines with $N_{\rm HI}$ determined from \Lya\ profile fits. 
Red points are the 26 stars lacking \Lya\ fits for \HI\  (Tables 2 and 4), where $N_{\rm HI}$ 
was scaled from $E(B-V)$.  The mean ratio for the 112 stars is 
$\langle N_H/E(B-V) \rangle = (6.07\pm1.01) \times10^{21}~{\rm cm}^{-2}~{\rm mag}^{-1}$.  
The mean and (rms) deviations are shown as horizontal lines and blue wash.  }  
\end{figure}

\clearpage


\begin{figure*}[!h]
\includegraphics[angle=0,scale=0.55]{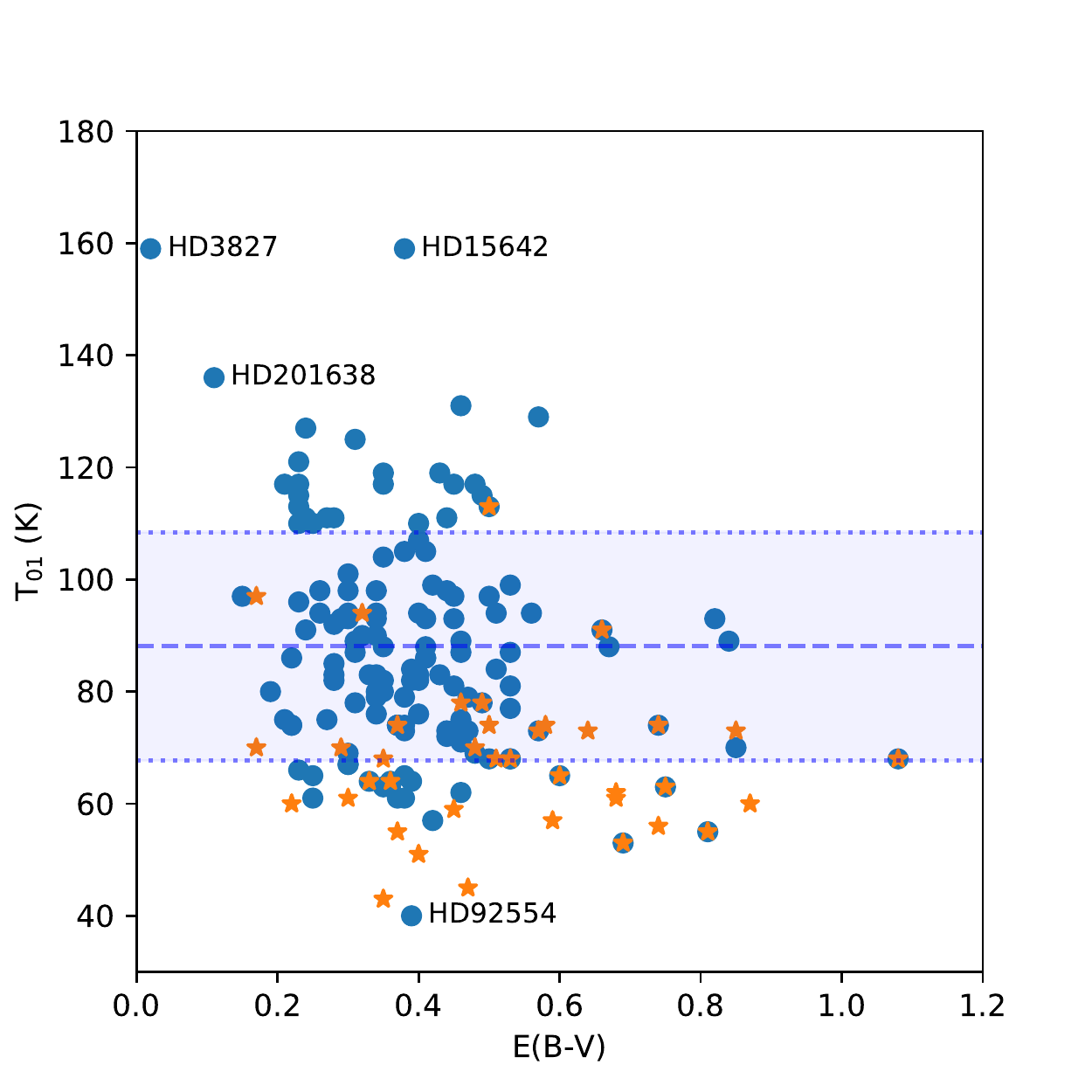}
\includegraphics[angle=0,scale=0.55]{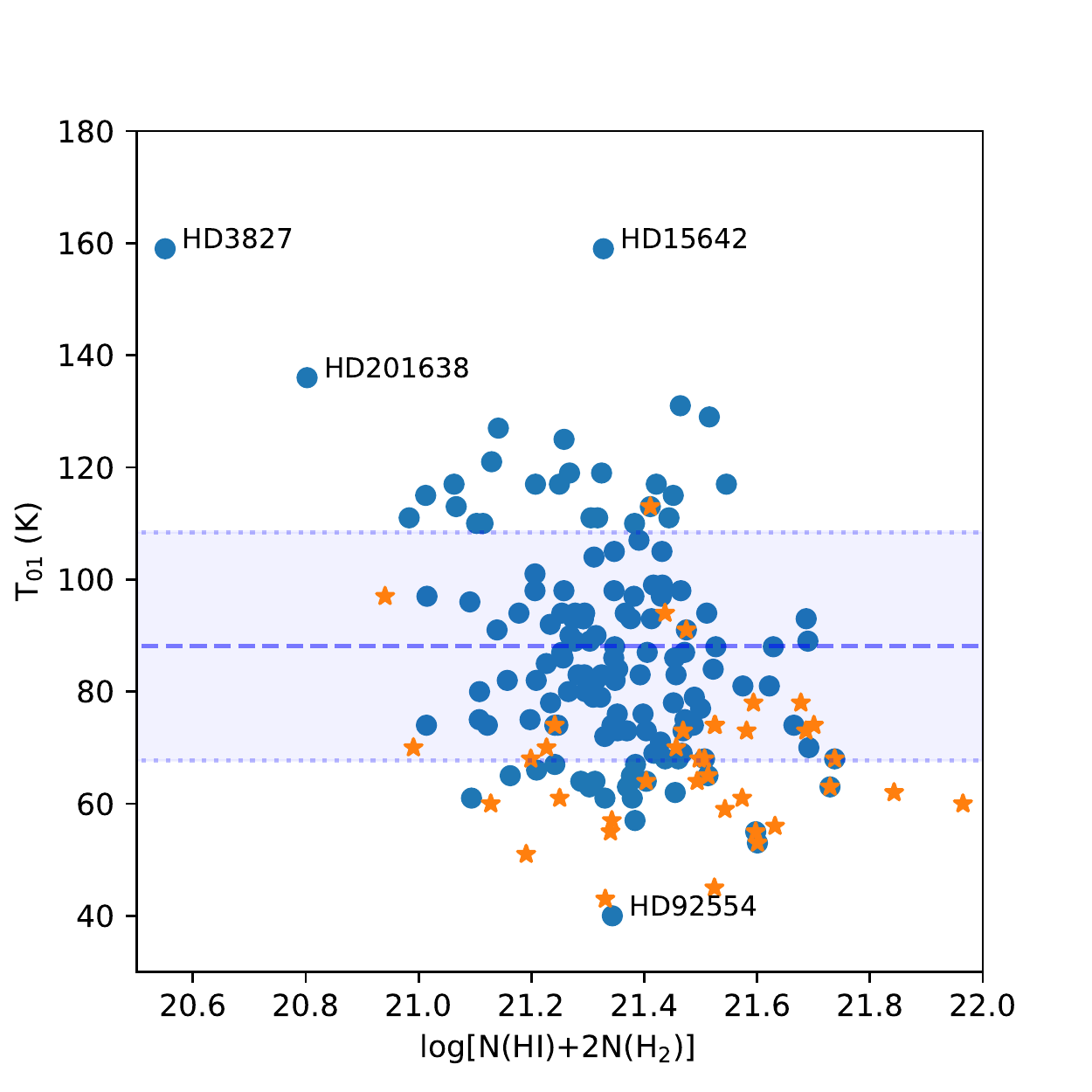}
\caption{Rotational temperature $T_{01}$ vs.\ color excess (left panel) and 
total hydrogen column density (right panel).  
The mean value $\langle T_{01} \rangle = 88 \pm 20$~K (horizontal dashed line 
with $1 \sigma$ dispersions) should track the gas kinetic temperature in high-density 
clouds. Lower temperatures appear in translucent clouds at $E(B-V) \gtrsim 0.5$ 
and $\log N_{\rm H} \gtrsim 21.5$.  Symbols are color-coded as in Figure 3 with four 
outlier targets labeled.  }
\end{figure*}


\begin{figure}[!h]
\includegraphics[angle=0,scale=0.55]{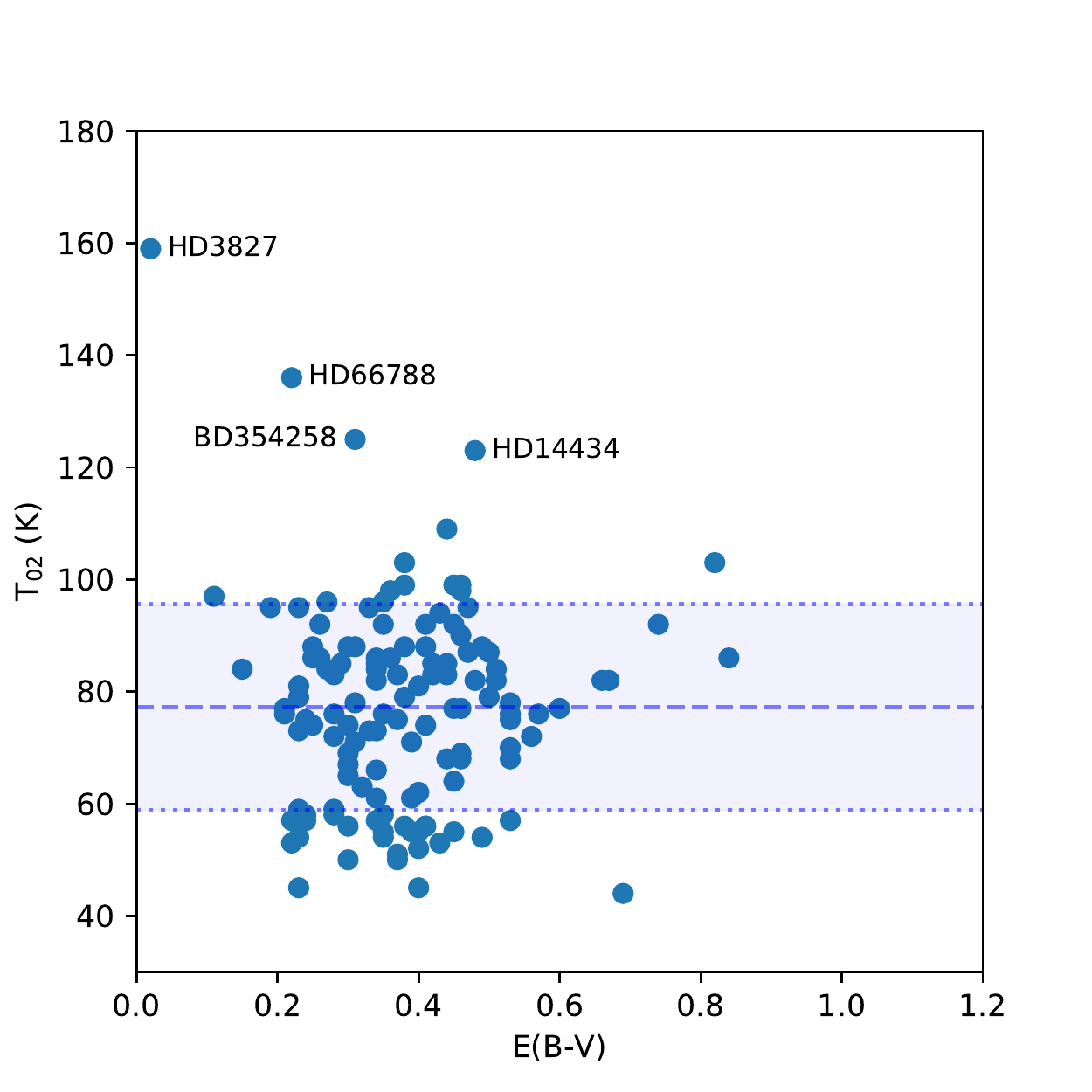}
\includegraphics[angle=0,scale=0.55]{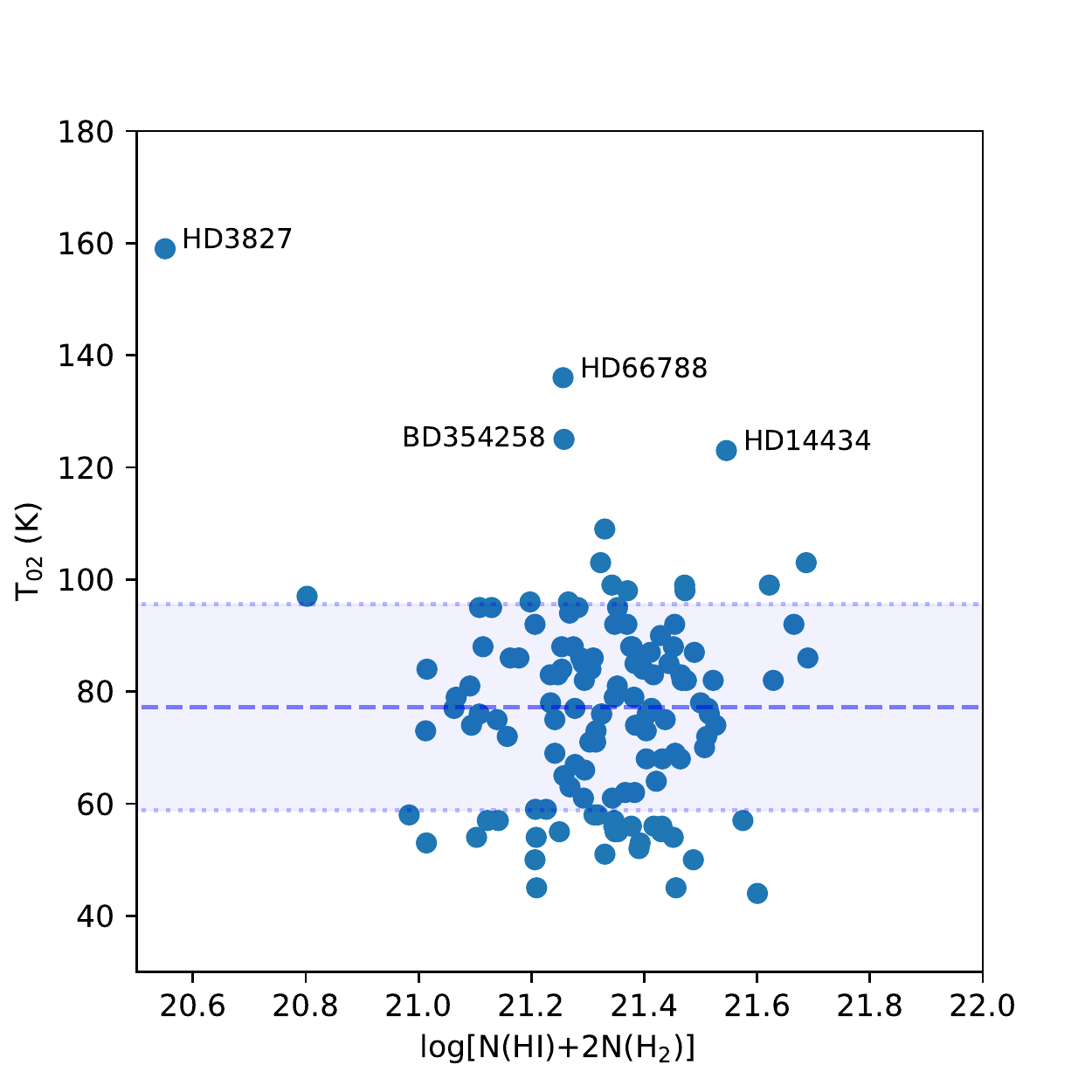}
\caption{Rotational temperature $T_{02}$ vs.\ color excess (left panel) and 
total hydrogen column density (right panel).  Four outlier targets are labeled.  
The other 128 stars have a mean value $\langle T_{02} \rangle = 77 \pm 18$~K  
(horizontal dashed lines with mean and $1 \sigma$ dispersions).  
Within individual sight-line errors (Section~3.4) and spreads of the distributions, 
$\langle T_{02} \rangle$ is similar to $\langle T_{01} \rangle = 88 \pm 20$~K.  
The lowest three rotational levels ($J = 0,1,2$) are likely coupled to the gas 
kinetic temperature. }
\end{figure}

\clearpage



\begin{figure}[!htb]
\includegraphics[angle=0,scale=1.0]{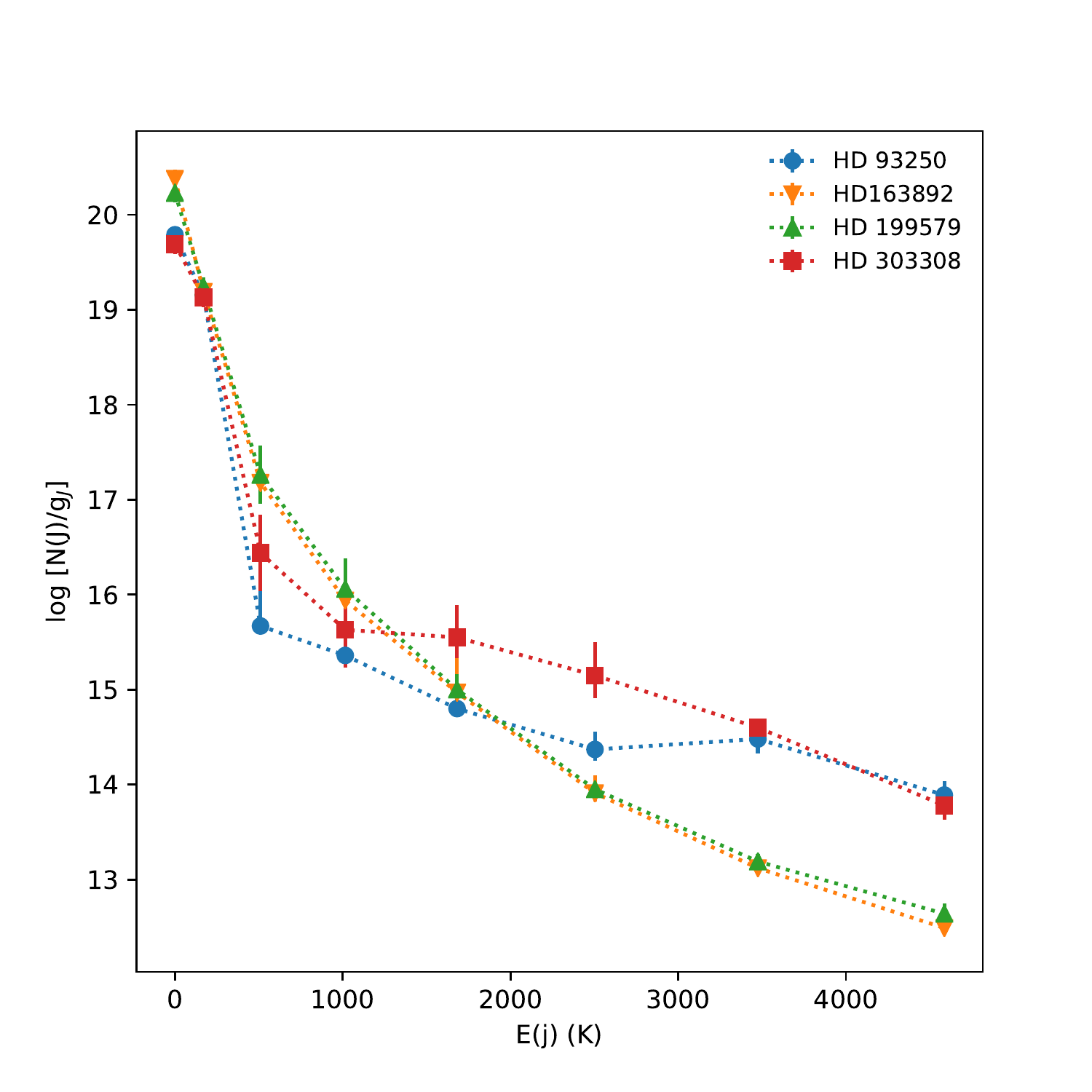}
\caption{Populations of \Htwo\ rotational states, $\log [N(J)/g_J]$, vs.\ their excitation 
energies, $E(J)/k$, expressed as temperatures.  The level statistical weights are
$g_J = (2S+1)(2J+1)$, where $S = 0$ (even-$J$) and $S = 1$ (odd-$J$). We show 
distributions ($J = 0-7$) for four sight lines with different excitation temperatures of 
low-$J$ and high-$J$ states.  The lowest levels ($J = 0,1,2$) appear thermally
coupled.  For HD~199579 (O6.5~V) we find $T_{01} = 74$~K and $T_{02} = 75$~K, 
but $T_{24} =  225$~K and $T_{35} = 307$~K.   For HD~163892 (O9~IV) we find 
$T_{01} = 62$~K and $T_{02} = 69$~K, but $T_{24} = 230$~K and $T_{35} = 319$~K.   
Two sight lines toward hotter stars show even higher excitation
($T_{24} = 582$~K, $T_{35} = 686$~K) for HD~93250 (O4~III) and 
($T_{24} = 569$~K, $T_{35} = 1358$~K) for HD~303308 (O4.5~V).  }
\end{figure}

\clearpage



\begin{figure}[!htb]
\includegraphics[angle=0,scale=0.65]{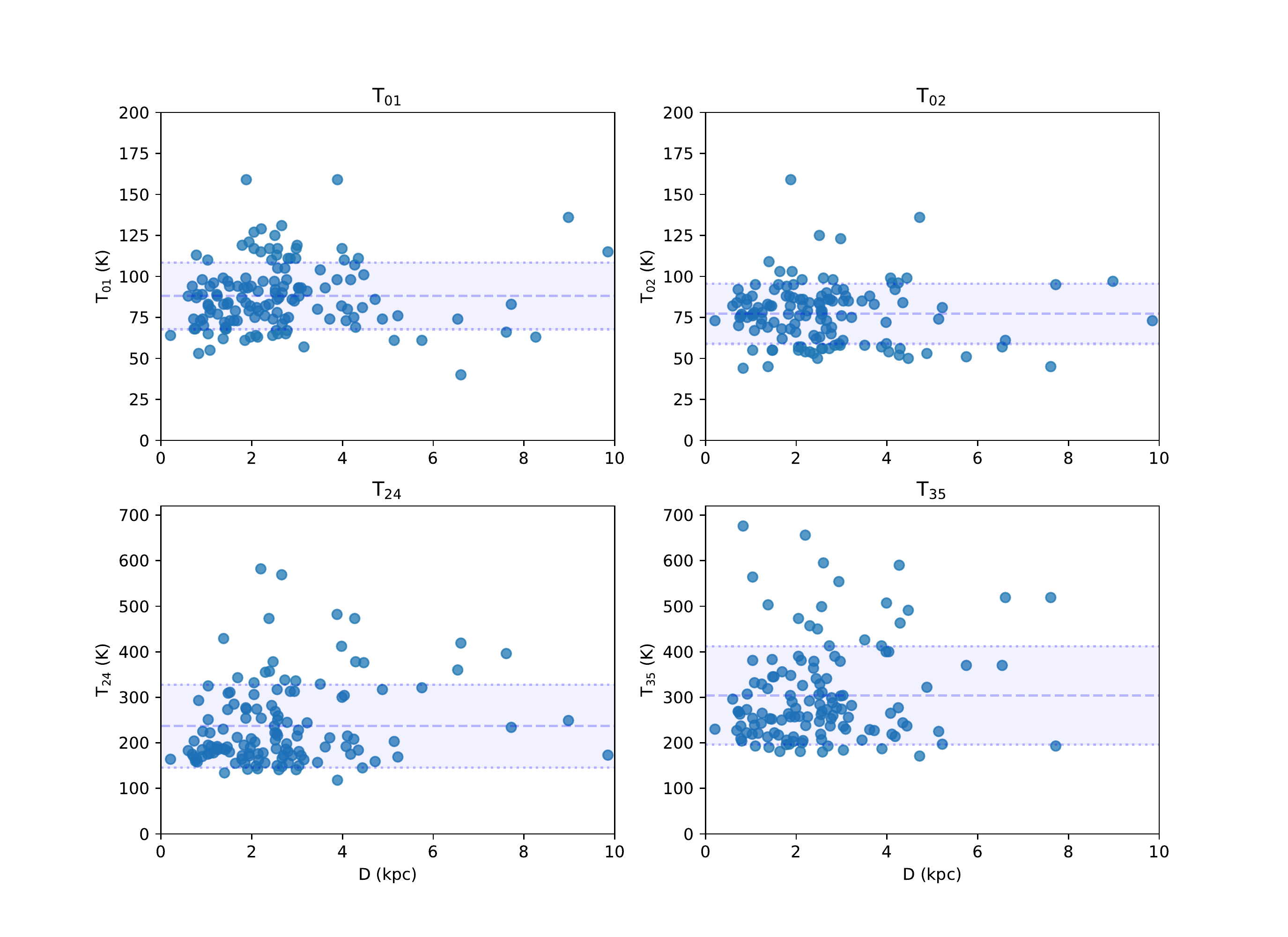}
\caption{Excitation temperatures, $T_{01}$ and $T_{02}$, of the lowest rotational
levels and $T_{24}$ and $T_{35}$ for higher excited levels, versus photometric 
distance to the target stars.  Several sight lines with missing or uncertain values of 
$T_{02}$, $T_{24}$,  $T_{35}$, have been omitted.  
Horizontal lines show means and $1 \sigma$ dispersions.  A significant fraction (10-15\%)
in the lower two panels have high values of  $T_{24}$ and $T_{35}$, lying above the 
(rms) dispersions. }
\end{figure}

\clearpage



\begin{figure}[!htb]
\includegraphics[angle=0,scale=1.0]{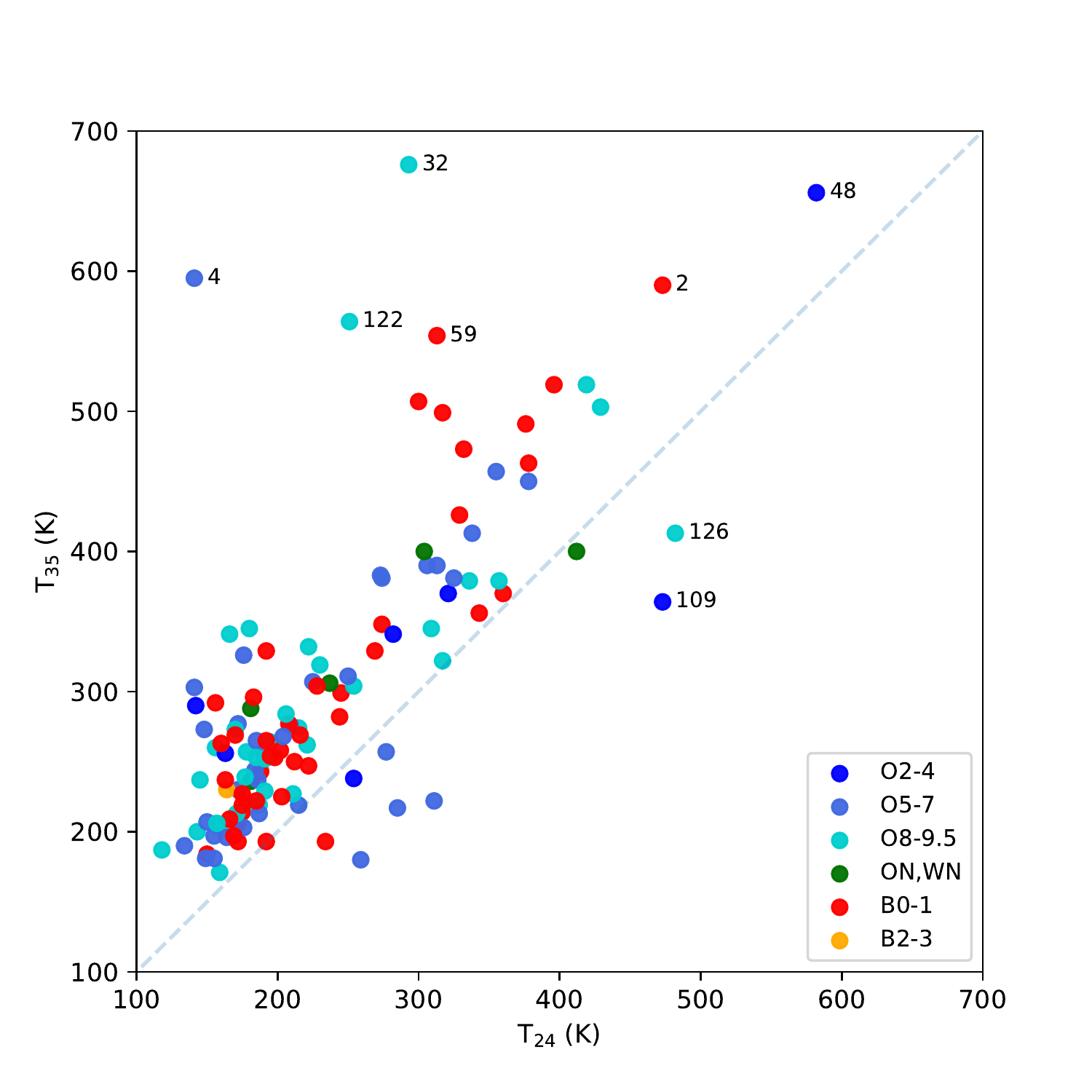}
\caption{Relation between pairwise rotational excitation temperatures, 
$T_{35}$ and $T_{24}$, connecting ortho (odd-$J$)  and para (even-$J$) states. 
These two temperatures are correlated above 300~K, usually with $T_{35} > T_{24}$.  
Most data points lie above the dashed line of slope unity.  
Stars of similar SpT are color-coded as follows:  
dark blue (O2--O4); cornflower blue (O5--O7); cyan (O8--O9.5); 
dark green (ON and WN); red (B0--B1); and orange (one B3 star).  
Using internal ID numbers, we label stars with high excitation temperatures, 
plus several that lie off the trend-line (see Section 3.5). 
Star \#137 (HD~303308, O4.5~Vfc) is not shown, with its off-scale excitation 
temperatures, $T_{35} = 1358$~K and $T_{24} = 569$~K.  }
\end{figure}

\clearpage


\clearpage

%
%


\begin{deluxetable}{llrrrrclllcr}
\tablecolumns{12}
\tabletypesize{\scriptsize}
\tablenum{1}
\tablewidth{0pt}
\tablecaption{Stellar Parameters\tablenotemark{a} and Distances\tablenotemark{b} }

\tablehead{
 \colhead{ID}
&\colhead{Target}
&\colhead{$\ell$}
&\colhead{$b$}
&\colhead{$B$}
&\colhead{$V$}
&\colhead{$E(B-V)$}
&\colhead{SpT}
&\colhead{$D_{\rm phot}$\tablenotemark{b}}
&\colhead{$D_{\rm Gaia}$\tablenotemark{b}} 
&\colhead{Program} 
&\colhead{$t_{\rm exp}$}
\\
 \colhead{}
& \colhead{}
& \colhead{(deg)}
& \colhead{(deg)}
& \colhead{(mag)}
& \colhead{(mag)}
&\colhead{(mag)}
&\colhead{}
&\colhead{(kpc)}
&\colhead{(kpc)}
&\colhead{(FUSE ID)}
&\colhead{(ksec)}
}

\startdata
1    & {BD $35^{\circ}4258$}    & 77.19   & -4.74   &  9.42 & 9.41  & 0.31 & B0.5\,Vn      & {\bf 2.51} & 1.89[1.71,2.10]  &P1017902 &  5.673   \\
2    & {BD $53^{\circ}2820$}    &101.24  & -1.69   &10.05 & 9.95  & 0.40 & B0\,IVn        & {\bf 4.27} & 3.18[2.82,3.65]  &P1223203 &  5.147   \\
3    & {CPD-$59^{\circ}2600$} &287.60  & -0.74   & 8.82  & 8.61  & 0.53 & O6\,Vf          & 2.11         & 5.11[4.28,6.36]  &P1221403 &  5.307   \\
4    & {CPD-$59^{\circ}2603$} &287.59  & -0.69   & 8.91  & 8.81  & 0.46 & O7.5\,V        & 2.60        & 3.66[3.22,4.22]  &S3040502  & 31.599  \\
5    & {CPD-$69^{\circ}1743$} &303.71  & -7.35   & 9.40  & 9.38  & 0.30 & B0.5\,IIIn     & {\bf 4.47} & 3.43[3.04,3.93]  &P1013701 &  3.548   \\
6    & {CPD-$72^{\circ}1184$} &299.15  &-10.94  &10.61 &10.68 & 0.23 & B0\,III          & {\bf 9.85} & 6.88[5.46,9.32]  &S5140103 &  9.172   \\
7    & {HD   3827}                       &120.79  &-23.23  & 7.76  & 8.01  & 0.02  & B0.7\,V       & {\bf 1.88} & 2.15[1.85,2.59]  &P1010302 &  2.885   \\
8    & {HD   5005A}                     &123.12  & -6.24   & 8.19  & 8.10 & 0.41 & O6.5\,Vf       &  2.89        & 2.71[2.39,3.14]  &P1020102  &  6.372   \\
\enddata
 
\tablenotetext{a} {Updated stellar parameters, photometry, spectral types, and distances are from
a survey of 139 OB-type stars (Shull \& Danforth 2019) with their internal target ID (column~1).  
These parameters are based on optical and near-IR digital photometry and extinction corrections
from the Galactic O-Star Spectroscopic Survey (Ma\'iz Apell\'aniz \etal\ 2004) and new spectral types 
(Sota \etal\ 2011, 2014).  Columns 3 and 4 list the Galactic longitude and latitude. 
Later columns list the stellar photometry [$B, V, E(B-V)$], spectral types (SpT),  photometric distances 
($D_{\rm phot}$), and parallax distances ($D_{\rm Gaia}$) listed in Tables 1 and 2 of 
Shull \& Danforth (2019).  The last two columns give the Program ID and exposure time of the
primary \FUSE\ spectroscopic observations. }
 
 \tablenotetext{b} {Photometric distances ($D_{\rm phot}$) are based on GOS photometry, extinctions, 
 and SpTs and a new set of absolute magnitudes (see Shull \& Danforth 2019).   When GOS data were
 not available, we based photometric distances (shown in {\bf boldface}) on photometry and SpT in the
 literature.  Parallax distances ($D_{\rm Gaia}$) and error ranges are based on parallaxes and errors 
 from the {\it Gaia}-DR2 archive, after adding a constant parallax offset of 0.03 mas.  }  
 
\end{deluxetable}



\begin{deluxetable}{rlcccccl}
\tablecolumns{8}
\tabletypesize{\scriptsize}
\tablenum{2}
\tablewidth{0pt}
\tablecaption{H~I Column Density Measurements\tablenotemark{a} }

\tablehead{
  \colhead{ID}
&\colhead{Target}
&\colhead{$E_{B-V}$\tablenotemark{b}}
&\colhead{$\log N_{\rm HI}$\tablenotemark{c} }
&\colhead{$\log N_{\rm HI}$}
&\colhead{$\log N_{\rm HI}$}
&\colhead{$\log N_{\rm HI}$}
&\colhead{$\log N_{\rm HI}$\tablenotemark{c} }
\\
 \colhead{}
& \colhead{(Star)}
& \colhead{(mag)}
& \colhead{(Adopted)}
& \colhead{(SVS85)}
& \colhead{(DS94)}
& \colhead{(J19)}
& \colhead{(Other)} 
}

\startdata
1    & {BD $35^{\circ}4258$}     & 0.31 & 21.24 &                                &                               & $21.24^{+0.03}_{-0.07}$ &       \\
2    & {BD $53^{\circ}2820$}     & 0.40 & 21.35 &                                &                               & $21.35^{+0.05}_{-0.07}$ &       \\
3    & {CPD -$59^{\circ}2600$} & 0.53 & 21.54 & $21.48\pm0.10$ & $21.54\pm0.07$ &                                              &  $21.48\pm0.15$ (FM90)  \\
4    & {CPD -$59^{\circ}2603$} & 0.46 & 21.43 & $21.48\pm0.15$ & $21.46\pm0.07$ & $21.43^{+0.04}_{-0.05}$ &       \\
5    & {CPD -$69^{\circ}1743$} & 0.30 & 21.16 &                               & $21.12 \pm0.11$ & $21.16^{+0.04}_{-0.09}$ &       \\     

6    & {CPD -$72^{\circ}1184$} & 0.23 & 20.90 &                                & $20.90\pm0.08$  &                                              &       \\    
7    & {HD   3827}                        & 0.02 & 20.55 &                                & $20.56\pm0.09$  & $20.55^{+0.07}_{-0.05}$ &       \\    
8    & {HD   5005A}                     & 0.41 & 21.40 & $21.48\pm0.15$ & $21.40\pm0.10$  &                                              &        \\
\enddata

\tablenotetext{a} {The adopted column densities $N_{\rm HI}$ (in cm$^{-2}$) in column 4 (SD20) 
are compared to three previous \Lya-fitting surveys in columns~5--7:  
SVS85 (Shull \& Van Steenberg 1985);  DS94 (Diplas \& Savage 1994);  and J19 (Jenkins 2019).  
Column 8 lists individual measurements from:  FM90 (Fitzpatrick \& Massa 1990);  
Rat02 (Rachford \etal\ 2002); Han92 (Hanson \etal\ 1992); Sno96 (Snow \etal\ 1996); and 
scaled estimates of $N_{\rm HI}$ from $E(B-V)$.  Footnote (c) to Table 3 provides further details. }

\tablenotetext{b} {Updated values of color excess  (Shull \& Danforth 2019) were derived using
digital photometry, visual extinction,  and new SpTs from the Galactic O-star Spectroscopic 
Survey (Ma\'iz Apell\'aniz \etal\ 2004; Sota \etal\ 2011, 2014). } 

\tablenotetext{c} {The adopted source for $N_{\rm HI}$ was selected in priority order of:  J19 (57 stars), 
DS94 (51 stars), FM90 (2~stars), and three other sources (3 stars).   For 26 stars that lack fits to \Lya\ 
profiles, we estimated $N_{\rm HI}$ by scaling with color excess $E(B-V)$;  see Section 2.1.}   

\end{deluxetable}



\pagebreak 
\begin{deluxetable}{rllll llllll}
\tablecolumns{10}
\tabletypesize{\scriptsize}
\tablenum{3}
\tablewidth{0pt} 
\tablecaption{Column Densities\tablenotemark{a} and Doppler Parameters\tablenotemark{a} }

\tablehead{
\colhead{ID} 
&\colhead{Target}
&\colhead{log N(0)}
&\colhead{log N(1)}
&\colhead{log N(2)}
&\colhead{log N(3)}
&\colhead{log N(4)} 
&\colhead{log N(5)}
&\colhead{log N(6)}
&\colhead{$b$ (\kms)}
}

 
\startdata
1     & {BD $35^{\circ}4258$} & 19.03$^{+0.10}_{-0.10}$ & 19.39$^{+0.10}_{-0.10}$ & 17.72$^{+0.21}_{-0.30}$  
                            & 17.78$^{+0.23}_{-0.32}$ & 15.68$^{+0.20}_{-0.14}$  & 15.36$^{+0.17}_{-0.12}$  
                            & 14.14$^{+0.18}_{-0.18}$ & 8.4$^{+0.9}_{-1.0}$  \\
2    & {BD $53^{\circ}2820$} & 19.59$^{+0.10}_{-0.10}$ & 19.85$^{+0.10}_{-0.10}$ & 16.07$^{+0.40}_{-0.40}$ 
                           & 15.85$^{+0.4}_{-0.4}$ & 15.25$^{+0.04}_{-0.02}$ &14.95$^{+0.03}_{-0.02}$  & $<14.15$ & 20$^{+1.0}_{-1.0}$  \\
3    & {CPD $-59^{\circ}2600$} & 19.84$^{+0.10}_{-0.10}$ & 19.88$^{+0.10}_{-0.10}$ & 16.66$^{+0.46}_{-0.32}$  
                           & 16.39$^{+0.36}_{-0.20}$  &15.06$^{+0.10}_{-0.07}$  
                           & 14.89$^{+0.08}_{-0.06}$  & 14.04$^{+0.10}_{-0.10}$  & 8.4$^{+0.9}_{-1.0}$    \\
4    & {CPD $-59^{\circ}2603$} & 19.77$^{+0.05}_{-0.05}$ & 19.87$^{+0.10}_{-0.10}$ & 18.23$^{+0.50}_{-0.50}$ 
                           & 15.65$^{+0.11}_{-0.73}$ & 14.88$^{+0.05}_{-0.05}$ & 14.76$^{+0.06}_{-0.06}$  
                           & $13.94^{+0.18}_{-0.18}$ & 11.3$^{+0.9}_{-1.0}$  \\
5    & {CPD $-69^{\circ}1743$}  & 19.49$^{+0.10}_{-0.10}$ & 19.71$^{+0.10}_{-0.10}$ & 15.78$^{+0.12}_{-0.08}$  
                           & 15.51$^{+0.10}_{-0.08}$ & 14.68$^{+0.09}_{-0.09}$  & 14.39$^{+0.04}_{-0.04}$ & $<13.85$  & 13.3$^{+0.9}_{-1.0}$  \\
6    & {CPD $-72^{\circ}1184$}  & 19.59$^{+0.10}_{-0.10}$ & 19.90$^{+0.10}_{-0.10}$ & 17.24$^{+0.70}_{-0.50}$  
                           & 16.04$^{+1.43}_{-0.20}$ & 14.56$^{+1.38}_{-0.03}$ & $<14.30$ & $<14.1$0 & \ldots    \\
7    & {HD   3827} & 16.84$^{+0.05}_{-0.05}$ & 17.33$^{+0.05}_{-0.05}$ & 16.15$^{+0.20}_{-0.13}$ & 15.82$^{+0.14}_{-0.10}$  
                           & 14.55$^{+0.06}_{-0.06}$ &14.16$^{+0.09}_{-0.09}$  & $<13.85$ & 12.8$^{+0.9}_{-1.0}$  \\
\enddata

\tablenotetext{a}{Column densities, $N(J)$ (in cm$^{-2}$), in rotational states $J$ derived from 
curve-of-growth fitting with Doppler parameter ($b$ in last column) and damping wings for lines 
from $J = 0$ and $J = 1$.  No value is listed when $b$ was poorly determined. Error bars on 
low-$J$ states ($J=  0, 1$ and sometimes $J = 2, 3$) depend on data quality (see footnote in 
Table~3).   In damping-wing fits, errors on $\log N$ typically range from $\pm 0.03$ to $\pm 0.10$.   
Six targets had detectable column densities in $J = 7$:   \#48 with $\log N(7)= 15.54\pm0.15$, 
\#52 with $\log N(7)= 14.47\pm0.05$, \#88 with $\log N(7)= 14.14\pm0.06$, \#116 with 
$\log N(7)= 14.29\pm0.11$, \#127 with $\log N(7)= 14.25\pm0.05$, and  \#137 with 
$\log N(7)= 15.43\pm0.15$ (15.32 in blue component and 14.80 in red component). }

\tablenotetext{b}{In 15 sight lines, multiple velocity components were observable and measured in
high-$J$ states.  We report individual column densities, denoted as  ``blue and red" components, and 
sum them to find total column densities.  } 

\tablenotetext{c}{A few targets have no listed column densities $N(2)$ or $N(3)$.  
In several cases this was a result of poor data quality, so that lines from $J \geq 2$ could not be 
fitted accurately.  In other cases, the accessible (unblocked) lines from $J = 2$ and $J = 3$ were 
highly saturated, with equivalent widths of 200-300~m\AA\ and large effective Doppler parameters 
($b  \geq10$~\kms) likely produced by unresolved velocity components.   When equivalent widths
of all the accessible lines lie on the  ``flat portion" of the curve of growth, errors can exceed 
$\pm0.40$ for $\log N$ = 16-17 and $\pm0.50$ for $\log N$ = 17-19.  }

\end{deluxetable}


%
%

\pagebreak 
\begin{deluxetable}{rll rrrr  llllll}
\tablecolumns{13}
\tabletypesize{\scriptsize}
\tablenum{4}
\tablewidth{0pt}
\tablecaption{Column Densities\tablenotemark{a} and Rotational Temperatures\tablenotemark{b} }

\tablehead{
\colhead{ID} 
&\colhead{Target}
&\colhead{$\log N_{\rm H2}$} 
&\colhead{T$_{01}$}
&\colhead{T$_{02}$}
&\colhead{T$_{\rm 24}$}
&\colhead{T$_{\rm 35}$}
&\colhead{$\log N_{\rm HI}$\tablenotemark{c} }
&\colhead{$\log N_{\rm H}$\tablenotemark{c} }
&\colhead{$D_{\rm phot}$\tablenotemark{d}} 
&\colhead{$\langle n_{\rm H} \rangle$\tablenotemark{d}}
&\colhead{$f_{\rm H2}$\tablenotemark{d} }
&\colhead{Notes\tablenotemark{a} }
\\
\colhead{}
&\colhead{}
&\colhead{}
&\colhead{(K)}
&\colhead{(K)}
&\colhead{(K)} 
&\colhead{(K)} 
&\colhead{}
&\colhead{}
&\colhead{(kpc)}
&\colhead{(cm$^{-3}$)}
&\colhead{}
&\colhead{}
}

\startdata
1    & {BD $35^{\circ}4258$}      & 19.56  & 125   & 125  & 222 & 247    & 21.24 & 21.26 & 2.51 & 0.22  & 0.040   & 2  \\
2    &  {BD $53^{\circ}2820$}     & 20.04  & 107   &   52  & 473 & 590    & 21.35 & 21.39 & 4.27 & 0.17  & 0.089  & 2 \\
3    & {CPD $-59^{\circ}2600$}  & 20.16  & 81     &   57  & 274 & 381    & 21.54 & 21.57 & 2.11 & 0.57  & 0.077  & 2  \\
4    & {CPD $-59^{\circ}2603$}  & 20.13  & 87     &   99  & 141 & 595    & 21.43 & 21.47 & 2.60 & 0.37  & 0.091  & 1  \\
5    & {CPD $-69^{\circ}1743$}  & 19.91  & 101   &  50   & 376 & 491    & 21.16 & 21.21 & 4.47 & 0.36  & 0.101  & 2  \\
6    & {CPD $-72^{\circ}1184$}  & 20.07  & 115   &  73   & 173 & \ldots & 20.90 & 21.01 & 9.85 & 0.034 & 0.228  & 2  \\
7    & {HD   3827}                         & 17.48  & 159   & 159  & 274 & 348    & 20.55 & 20.55 & 1.88 & 0.061 & 0.0017 & 1  \\
8    & {HD   5005A}                      & 20.22  & 86     &   92   & 172 & 277   & 21.40 & 21.45 & 2.89 & 1.03  & 0.117   &  1  \\
\enddata
\end{deluxetable}



\begin{deluxetable}{llccc}
\tablecolumns{5}
\tabletypesize{\footnotesize}
\tablenum{5}
\tablewidth{0pt}
\tablecaption{Observational Averages from \FUSE\ Survey\tablenotemark{a} }

\tablehead{
   \colhead{Quantity}
& \colhead{Units}
& \colhead{$D\leq 2$~kpc}
& \colhead{$D\leq 5$~kpc}
& \colhead{All}   
}

\startdata
Sample                                                  &    \# Stars                           &   56       & 129        &  138   \\
$\langle D \rangle$                             &   kpc                                    &   1.31    & 2.27      &  2.62   \\
$\langle n_{\rm HI} \rangle$              &   cm$^{-3}$                        &   0.553  & 0.364   &  0.344  \\
$\langle n_{\rm H2} \rangle$             &   cm$^{-3}$                       &   0.261  & 0.136   &  0.129   \\
$\langle n_{\rm H} \rangle$               &  cm$^{-3}$                         &   0.814  & 0.500   &  0.472  \\
$\langle f_{\rm H2} \rangle$              &  $\dots$                             &   0.268   & 0.202   &  0.200  \\
$\langle E(B-V)/D \rangle$                &  mag~kpc$^{-1}$             &  0.434  & 0.260     & 0.246   \\
\enddata

\tablenotetext{a} {Mean values of total hydrogen density, $n_{\rm H} = N_{\rm H}/D_{\rm phot}$, 
and molecular fraction $f_{\rm H2}$ along the sightlines for stars in the full  \FUSE\ survey 
and for sub-samples within two photometric distance limits  (Shull \& Danforth 2019).  The 
averages are column density-weighted (see Section 3.1).   
Column~3 is for 56 stars with $D_{\rm phot} \leq 2$~kpc,
Column~4 is for 129 stars with $D_{\rm phot} \leq 5$~kpc,
and Column 5 is for the full sample, omitting HD~3827 (uncertain $E(B-V)$.
Statistics for $N_H/E(B-V)$ omit 26 stars without \Lya\ profile fits for \HI. 
 }   

\end{deluxetable}

\end{document}